\documentclass{article}

\usepackage{microtype}
\usepackage{graphicx}
\usepackage{multirow}
\usepackage{subcaption}
\usepackage{booktabs} % for professional tables

\usepackage{hyperref}

 \usepackage[accepted]{icml2026}

\usepackage{amsmath}
\usepackage{amssymb}
\usepackage{mathtools}
\usepackage{amsthm}

\usepackage[capitalize,noabbrev]{cleveref}

\theoremstyle{plain}

\theoremstyle{definition}

\theoremstyle{remark}

\usepackage[textsize=tiny]{todonotes}

\icmltitlerunning{Discontinuous Galerkin Neural Operator for Pathology Defocus Deblurring}

\begin{document}

\twocolumn[
  \icmltitle{Discontinuous Galerkin Neural Operator for Pathology Defocus Deblurring}

  % It is OKAY to include author information, even for blind submissions: the
  % style file will automatically remove it for you unless you've provided
  % the [accepted] option to the icml2026 package.

  % List of affiliations: The first argument should be a (short) identifier you
  % will use later to specify author affiliations Academic affiliations
  % should list Department, University, City, Region, Country Industry
  % affiliations should list Company, City, Region, Country

  % You can specify symbols, otherwise they are numbered in order. Ideally, you
  % should not use this facility. Affiliations will be numbered in order of
  % appearance and this is the preferred way.
  \icmlsetsymbol{equal}{*}

  \begin{icmlauthorlist}
    \icmlauthor{Shaoqing Duan}{yyy}
    \icmlauthor{Haofei Song}{yyy}
    \icmlauthor{Xintian Mao}{yyy}
    \icmlauthor{Qingli Li}{yyy}
    \icmlauthor{Yan Wang}{yyy}
%    \icmlauthor{Firstname6 Lastname6}{sch,yyy,comp}
%    \icmlauthor{Firstname7 Lastname7}{comp}
    %\icmlauthor{}{sch}
%    \icmlauthor{Firstname8 Lastname8}{sch}
%    \icmlauthor{Firstname8 Lastname8}{yyy,comp}
    %\icmlauthor{}{sch}
    %\icmlauthor{}{sch}
  \end{icmlauthorlist}

%  \icmlaffiliation{yyy}{Department of XXX, University of YYY, Location, Country}
%  \icmlaffiliation{comp}{Company Name, Location, Country}
%  \icmlaffiliation{sch}{School of ZZZ, Institute of WWW, Location, Country}
  
  \icmlaffiliation{yyy}{Shanghai Key Laboratory of Multidimensional Information Processing, East China Normal University, Shanghai, China}

%  \icmlcorrespondingauthor{Firstname1 Lastname1}{first1.last1@xxx.edu}
  \icmlcorrespondingauthor{Yan Wang}{ywang@cee.ecnu.edu.cn}

  % You may provide any keywords that you find helpful for describing your
  % paper; these are used to populate the "keywords" metadata in the PDF but
  % will not be shown in the document
  \icmlkeywords{Machine Learning, ICML}

  \vskip 0.3in
]

% this must go after the closing bracket ] following \twocolumn[ ...

% This command actually creates the footnote in the first column listing the
% affiliations and the copyright notice. The command takes one argument, which
% is text to display at the start of the footnote. The \icmlEqualContribution
% command is standard text for equal contribution. Remove it (just {}) if you
% do not need this facility.

% Use ONE of the following lines. DO NOT remove the command.
% If you have no special notice, KEEP empty braces:
\printAffiliationsAndNotice{}  % no special notice (required even if empty)
% Or, if applicable, use the standard equal contribution text:
% \printAffiliationsAndNotice{\icmlEqualContribution}

% \twocolumn[{%
%	\renewcommand\twocolumn[1][]{#1}%
%	\maketitle
%	\begin{center}
%		\centering
%		\captionsetup{type=figure}
%		\vspace{-1.5em}    \includegraphics[width=0.96\linewidth]{fig/fig5_1.pdf}
%		\vspace{-1.em}
%		\caption{
%			\label{fig:f1}
%			Throughput / GPU memory / FLOPs comparisons of our proposed approach (vHeat-SQ) with SwinIR, Restormer, MPT under different image resolutions. The throughput and GPU memory are tested on 80 GB A100 GPUs with batch size 1. } 
%		% SwinIR is tested with scaled window size here.
%	\end{center}%
%	% \vspace{-1.em}
%}]

\begin{abstract}

  Defocus deblurring in pathological microscopy remains challenging due to the spatially varying and locally discontinuous nature of optical blur induced by a position-dependent integral imaging process.
  Existing deep learning methods, constrained by shift-invariance assumptions and limited interpretability, are not well suited to such heterogeneous blur patterns.
  Neural operators provide a principled alternative by modeling defocus formation directly as an integral operator, offering a new perspective on defocus deblurring.
  However, most existing neural operator architectures for low-level vision rely on globally parameterized kernels that assume smoothness and stationarity, limiting their ability to model heterogeneous and locally discontinuous blur patterns.
  To address this limitation, we propose the Discontinuous Galerkin Neural Operator (DGNO), which parameterizes the integral kernel using a discontinuous Galerkin formulation with element-local volume operators and interface numerical fluxes.
  DGNO provides a principled combination of locality,
  heterogeneity modeling, and global coherence while preserving the underlying physics
  of optical image formation. 
  Extensive and insightful experiments demonstrate that DGNO surpasses state-of-the-arts, delivering sharper reconstructions, robust handling of spatially varying blur, and scalable high-resolution performance. The code will be released at https://github.com/DeepMed-Lab-ECNU/Single-Image-Deblur.
\end{abstract}

\section{Introduction}

\begin{figure}[t]
% \vspace{-10pt}
	\centering
	\begin{center}
		\includegraphics[width=0.48\textwidth]{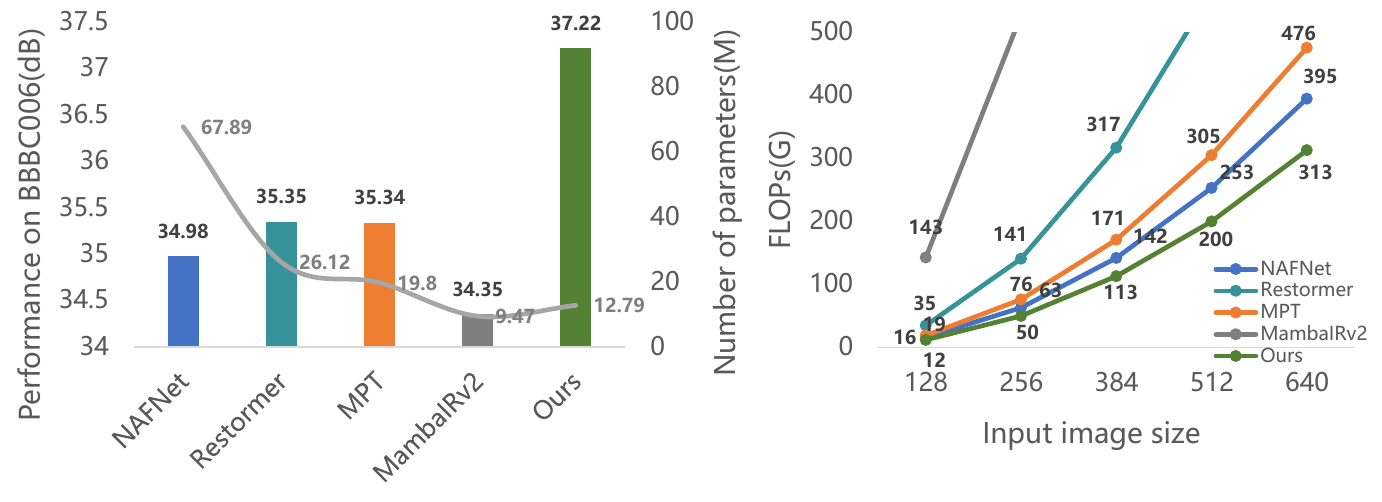}
        \vspace{-2em}
		\caption{
			Comparisons Our DGNO and other state-of-the-art algorithms. Performance and parameters on BBBC006$_{w1}$ (left) and FLOPs (right) for defocus deblurring.
		}
		\label{fig:performance}
	\end{center}
%	\vspace{-20pt}
	\vspace{-2em}
\end{figure}

Defocus deblurring for pathological microscopic images is of critical importance,
as optical defocus can severely degrade cellular morphology and compromise
downstream pathological analysis~\cite{zhang2022benchmarking}, including cell detection~\cite{schmidt2018cell} and segmentation~\cite{keaton2023celltranspose}. 
Many image defocus deblurring networks have been proposed.
However, when these methods are directly applied to
pathological defocus removal, their performance is often far from satisfactory,
as illustrated in Fig.~\ref{fig:performance} (see NAFNet~\cite{chen2022simple}, Restormer~\cite{zamir2022restormer}, MPT~\cite{zhang2024unified}, and MambaIRv2~\cite{guo2025mambairv2}). 
This is because most existing defocus deblurring methods aim to learn a finite-dimensional mapping
$\mathcal{F}:\;
\mathbb{R}^{N} \;\longrightarrow\; \mathbb{R}^{N}$,
where both the blurred and sharp images are represented as discrete pixel vectors defined on a fixed sampling grid~\cite{nah2017deep, tao2018scale}, without any support from physical imaging process.
But in the real world, physically faithful defocus deblurring process corresponds to a function-to-function mapping
$\mathcal{G}:\;
g(\cdot)\;\longmapsto\; h(\cdot),$
which models the continuous image formation process over the spatial domain~\cite{kovachki2023neural, li2020fourier, lu2021learning}. To the best of our knowledge, this perspective has rarely been exploited.

\begin{figure}[t]
	\centering
	\begin{center}
		\includegraphics[width=0.48\textwidth]{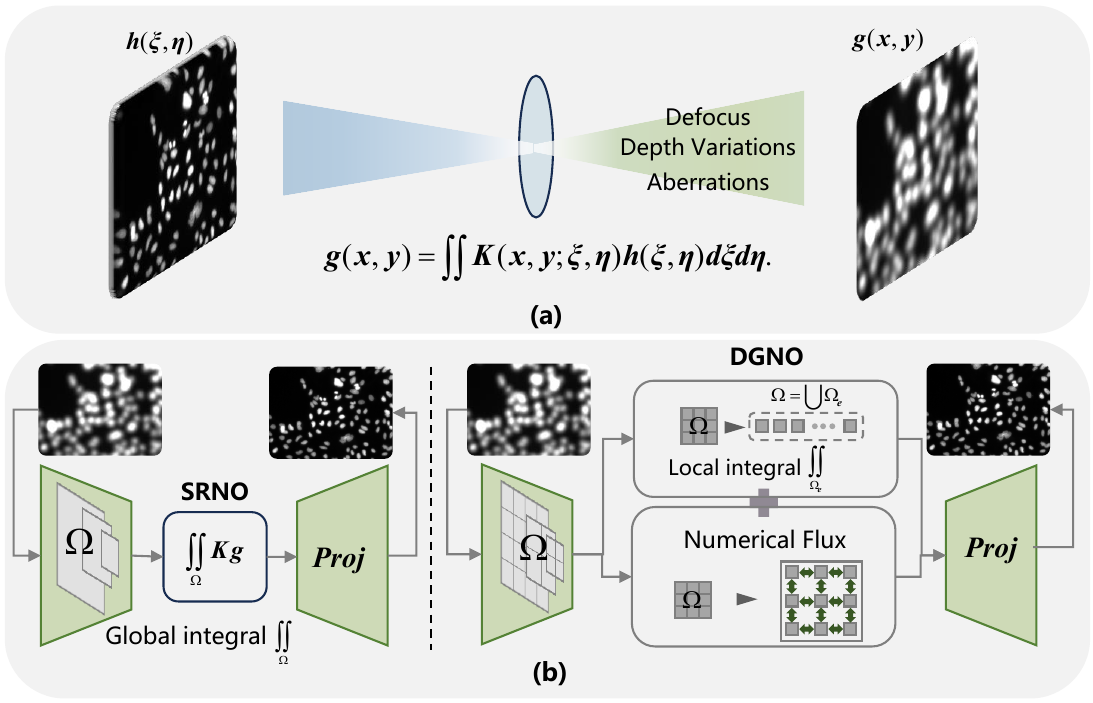}
        \vspace{-1.6em}
		\caption{
			Overview of defocus-blur formation and operator-learning approaches.
			(a) Defocus-Blur Formation.
			(b) SRNO versus the proposed DGNO.
		}
		\label{fig:motivate}
	\end{center}
	\vspace{-23pt}
\end{figure}

To understand the origin of pathological defocus blur, we consider the physical image
formation process. 
When light from an object point $(\xi,\eta)$ propagates through a lens under imperfect focus, it spreads on the image plane according to a position-dependent point-spread function (PSF), as shown in Fig.~\ref{fig:motivate} (a).
In practice, the defocus PSF is often well-approximated by a Gaussian-like disk kernel whose scale varies with the local degree of defocus~\cite{quan2021gaussian, quan2024deep}.
 As a result, the intensity at each image location $(x,y)$ is generated by an aggregation of contributions from a neighborhood region of the object domain. 
 Classical Fourier optics~\cite{goodman2005introduction} formalizes this process through a spatially varying integral operator:
\begin{equation}
	g(x,y) = \iint K(x,y;\xi,\eta)\, h(\xi,\eta)\, d\xi d\eta,
	\label{eq:defocus_integral}
\end{equation}
%where $K(x,y;\xi,\eta) = h(x,y;\xi,\eta)$
where $K$ denotes the PSF determined by the optical system. 
%Only under the restrictive assumption that the PSF is shift-invariant does the 
%Eq.~\eqref{eq:defocus_integral} reduce to a standard convolution:
%\begin{equation}
%	g(x,y)
%	= \iint h(x-\xi,y-\eta)f(\xi,\eta)d\xi d\eta
%	= (h * f)(x,y).
%\end{equation}
Only under the restrictive assumption of shift invariance does this formulation reduce
to a standard convolution.
But, this assumption is rarely satisfied in pathological microscopic imaging. 
Depth variations, heterogeneous tissue structures, refractive index inhomogeneity, and
spatially varying aberrations cause the PSF to vary across the field of view, making convolution inadequate. 
Besides, pathological microscopic images are characterized by piecewise structural heterogeneity, where local regions remain internally stable but exhibit consistent inter-region transitions. %As illustrated in Fig.~\ref{fig:motivate}(b), this behavior can be reflected by consistently lower coefficients of variation (CV) of gradient-energy jumps (Appendix~\ref{app:gradE_metrics}) for pathology images (0.52), compared to the substantially higher variability observed in natural images (0.73), indicating more stable and repeatable spatial variations rather than scene-dependent fluctuations.
Thus, pathology defocus deblurring should be formulated as an inverse problem of the
spatially varying integral operator in Eq.~\eqref{eq:defocus_integral}, where the goal is
to recover the latent sharp image $h$ from the observed blurred image $g$. This inverse
mapping is severely ill-posed due to the position-dependent and locally discontinuous nature of the PSF.

%(2) Pathological microscopic images are characterized by piecewise structural heterogeneity, where local regions remain internally stable but exhibit consistent inter-region transitions. As illustrated in Fig.~\ref{fig:motivate}(a), this behavior is reflected by consistently lower coefficients of variation (CV) of gradient-energy jumps (Appendix.\ref{app:gradE_metrics}) for pathology images (0.52), compared to the substantially higher variability observed in natural images (0.73), indicating more stable and repeatable spatial variations rather than scene-dependent fluctuations.

These challenges motivate learning-based approaches that directly approximate the
inverse mapping of the defocus operator from data, rather than relying on explicit
kernel estimation or convolutional assumptions.
Following this paradigm, a wide range of deep learning models have been applied to
defocus deblurring, including convolutional neural networks (CNNs), vision
transformers (ViTs), and more recent state-space models such as Mamba.
CNN-based methods~\cite{quan2021gaussian,cho2021rethinking, chen2022simple, quan2024deep} implicitly assume shift-invariant convolution and therefore struggle
to model spatially varying blur, while transformer-based approaches rely on global
self-attention to capture long-range dependencies without physical interpretability~\cite{liu_swin_2021, zamir2022restormer, zhang2024unified}.
Recent state-space models, such as Mamba~\cite{guo2024mambair, guo2025mambairv2}, further improve computational efficiency by
reducing complexity to linear time; however, they remain physically unstructured and
operate at the feature-sequence level, making them insufficient for modeling the
spatially varying integral operators that govern defocus blur.
Despite their architectural differences, these approaches fundamentally
treat defocus deblurring as a finite-dimensional image regression problem, rather than
as the inversion of a spatially varying integral operator.

From this perspective, defocus deblurring is more appropriately viewed as an operator
learning task, where the goal is to approximate the inverse of a spatially varying
integral operator acting on \textbf{function spaces}.
Recently, Neural Operators (NO)~\cite{li2020fourier,kovachki2023neural} have emerged as a powerful framework for learning
mappings between infinite-dimensional function spaces and directly parameterizing
integral operators, naturally aligning with this formulation.
Nevertheless, most existing NO used in low-level vision depend on global kernel parameterizations, including Diffusion Fourier Neural Operators (DiffFNO)~\cite{liu2025difffno} and Super-Resolution Neural Operator (SRNO)~\cite{wei2023super}.
These approaches implicitly assume smoothness and stationarity, making them unsuitable for
the highly localized and spatially heterogeneous behavior of real defocus blur. 
These motivate a defocus deblurring framework that retains the advantages
of NO while explicitly incorporating locality, heterogeneity, and
discontinuity awareness, which are essential for restoring images degraded by real
optical blur.

To this end,  we propose the \textbf{Discontinuous Galerkin Neural Operator (DGNO)}, a new operator-learning framework inspired by the discontinuous Galerkin (DG) method~\cite{hesthaven2008nodal}. 
As illustrated in Fig.~\ref{fig:motivate}(b), DGNO parameterizes the integral kernel in a
DG-style manner by decomposing the global integral kernel into element-local operators
and interface numerical fluxes.
The former explicitly models the spatial locality inherent to real defocus blur, while
the latter enables controlled cross-element information exchange at element interfaces
without oversmoothing local structures.
Moreover, DGNO supports both general face-based numerical flux formulations and a
lightweight zero-order DG (P0DG) approximation, allowing interface coupling to be
constructed either from face-wise operators or directly from element-local volume
operators.
By unifying element-wise operator learning with flux-based interface coupling, DGNO
achieves a principled balance between local adaptability and global consistency, enabling
effective modeling of spatially heterogeneous and locally discontinuous defocus blur
while preserving coherent global restoration behavior beyond globally parameterized
neural operators.
Our contributions can be summarized as follows:

\vspace{-1em}

\begin{itemize}
	\item We present the first neural operator formulation of defocus deblurring by grounding the model in the fact that defocus blur arises from a spatially varying, locally supported integral operator, making neural operators a physically aligned alternative to other architectures.
    
	\item Building on this perspective, we propose the Discontinuous Galerkin Neural Operator (DGNO), which decomposes the global integral operator into element-local volume integral operators and interface numerical fluxes, including both a general interface-based flux formulation and a lightweight Zero-Order DG (P0DG) approximation for flexible cross-element coupling.
	
	\item Extensive experiments demonstrate that DGNO effectively captures spatially varying and discontinuous defocus blur, achieving superior restoration quality compared to state-of-the-art neural operator and image restoration approaches.
\end{itemize}

\section{Related Work}

\paragraph{Defocus Deblur.}
Defocus deblurring aims to restore images degraded by spatially varying defocus blur caused by optical defocusing.
Conventional methods typically follow a two-stage paradigm that first
estimates a defocus map~\cite{shi2015just,karaali2017edge,zhao2019defocus} and then applies non-blind deconvolution using
hand-crafted blur kernels~\cite{yuan2008progressive, ren2018deep, nan2020deep}, which often suffers from inaccurate kernel
modeling and ringing artifacts~\cite{yuan2007image}.
Recent deep learning approaches adopt end-to-end CNN-, Transformer-, or state-space-model-based architectures to handle spatially varying blur~\cite{cho2021rethinking, quan2021gaussian, liu_swin_2021, zamir2022restormer,zhang2024unified,guo2025mambairv2}, but they tend to favor either local processing or global modeling, 
relying on implicit shift-invariance assumptions or lacking physical interpretability.

\paragraph{Neural Operators.}
Neural operators have recently emerged as a powerful framework for learning mappings between infinite-dimensional function spaces, enabling discretization invariant solutions of partial differential equations~\cite{li2020fourier, lu2021learning, kovachki2023neural}.
Fourier Neural Operators (FNOs) parameterize operators through global spectral representations, demonstrating strong capability in modeling long-range dependencies and generalizing across resolutions.
In low-level vision,~\cite{wei2023super} extend neural operators with kernel-based Galerkin-type attention to approximate integral operators, enabling resolution-invariant super-resolution via dynamic latent basis learning, while~\cite{liu2025difffno} enhance neural operators with spectral representations and hybrid spatial–frequency fusion mechanisms to better preserve high-frequency details in arbitrary-resolution image reconstruction.
Despite these advances, most existing neural operator formulations
implicitly assume global continuity of the underlying function and rely
on globally coupled representations, which are suboptimal for problems
characterized by spatially varying or piecewise behaviors.
In contrast, our work introduces a discontinuous Galerkin neural
operator that decomposes the global operator into
element-local volume operators and interface fluxes, providing a
structured mechanism to balance localized modeling and global coupling, and offering a new perspective for defocus deblurring.

\section{Method}
\label{sec:method}

\begin{figure*}[!htb]
	\centering
	\begin{center}
		\includegraphics[width=0.85\textwidth]{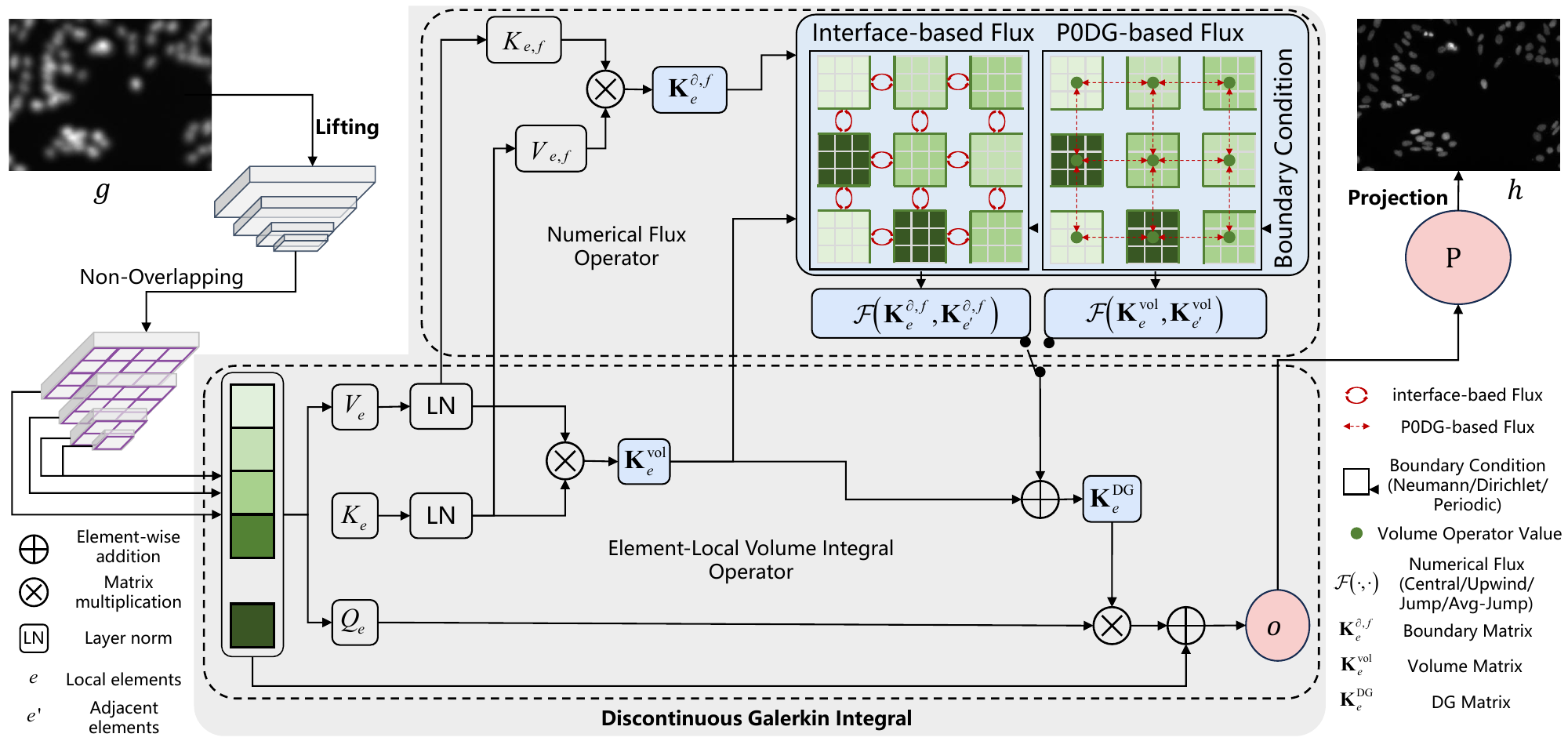}
        \vspace{-0.5em}
		\caption{
			The proposed Discontinuous Galerkin Neural Operator (DGNO) architecture for defocus deblurring by lifting a defocus image $x(r)$ into a feature space using a mamba encoder. Kernel integrals composed of $T$ layers of discontinuous attention. This pipeline generates $s(r)$, a sharp images of the input image.}
		\label{fig:DGNO_arch}
	\end{center}
	\vspace{-2.3em}
\end{figure*}

In this section, we first introduce the formulation of DGNO.
Then the parameterized integral operators are presented, including the element-local volume integral operator and the interface flux operator under both the interface-based DG formulation and its P0DG approximation.
Finally, we present the discrete operator assembly and network details.

\subsection{Neural Operator Preliminaries}
\label{sec:no_prelim}
We consider the problem of learning mappings between function spaces.
Given an input function $a : D \rightarrow \mathbb{R}^{d_a}$, a neural
operator aims to learn an operator $\mathcal{G}$ such that
$u(\cdot) = \mathcal{G}(a(\cdot))$.
Following the neural operator framework, the input function is first
lifted to a higher-dimensional latent representation by a pointwise
mapping $z_0(x) = P(a(x))$, where $P$ is applied independently at each
location.
The latent feature field is then updated through a sequence of operator
layers $z_0 \mapsto z_1 \mapsto \cdots \mapsto z_T$,
and the final output is obtained via a pointwise projection
$u(x) = Q(z_T(x))$.
Each operator layer follows a residual form consisting of a local linear
term and a non-local integral operator
\begin{equation}
	z_{t+1}(x)
	=
	\sigma\!\left(
	W z_t(x)
	+
	(\mathcal{K} z_t)(x)
	\right),
	\label{eq:no_update}
\end{equation}
where $\sigma$ denotes a pointwise nonlinearity, $W$ is linear operator.
The non-local operator $\mathcal{K}$ is defined as a kernel integral
operator acting on the entire domain
\begin{equation}
	(\mathcal{K} z)(x)
	=
	\int_D \kappa_\phi(x,y)\, z(y)\, dy,
	\label{eq:no_kernel}
\end{equation}
with $\kappa_\phi(x,y)$ denoting a learnable kernel function
parameterized by $\phi$.
Eq.~\eqref{eq:no_update} --~\eqref{eq:no_kernel} define a continuous
function-to-function mapping and serve as the strong-form representation
of neural operators.

\subsection{Discontinuous Galerkin Neural Operator}

\paragraph{DGNO Formulation.}
In discontinuous Galerkin (DG) methods for solving partial differential equations,
\textbf{integration by parts decomposes divergence-form differential operators
	into element-local volume terms and interface surface terms}, where
cross-element interactions are mediated by numerical fluxes on element
boundaries (Appendix~\ref{sec:dg_weak_form}).
Inspired by this volume--interface decomposition, DGNO introduces an
operator-level DG structure into the integral formulation of neural operators
in Eq.~\eqref{eq:no_kernel}.
Specifically, the integral term in Eq.~\eqref{eq:no_kernel} is a parameterized
kernel integral operator and does not contain differential terms, so it cannot
be directly decomposed by integration by parts.
Instead, DGNO decomposes the global kernel integral operator into element-wise
local operators and interface coupling operators.
The local integral part follows the element-wise neural operator construction,
while the interface part is modeled using classical numerical fluxes.

Following the neural operator update Eq.~\eqref{eq:no_update}, the DGNO iterative update is defined
globally on the domain $D$ as
\begin{equation}
	z_{t+1}(x)
	=
	\sigma\!\left(
	W z_t(x)
	+
	(\mathcal{K}^{\mathrm{DG}} z_t)(x)
	\right),
	\quad x \in D,
	\label{eq:dgno_update_global}
\end{equation}
where $\mathcal{K}^{\mathrm{DG}}$ denotes the DG neural operator. 
Given a partition of the spatial domain into non-overlapping elements
$D=\bigcup_{e=1}^{E}D_e$, the global DG operator is assembled from
element-local operators as
\begin{equation}
\small
	{(\mathcal{K}^{\mathrm{DG}} z)(x)
	=
	(\mathcal{K}^{\mathrm{DG}}_{e} z)(x),
	\quad x \in D_e,\; e=1,\dots,E,}
	\label{eq:dgno_operator_piecewise}
\end{equation}
where $\mathcal{K}^{\mathrm{DG}}_{e}$ denotes the element-local DG neural operator on element $D_e$. 
Accordingly, the element-local DG neural operator is written as
\begin{equation}
\small
	(\mathcal{K}^{\mathrm{DG}}_{e} z)(x)
	=
	(\mathcal{K}^{\mathrm{vol}}_{e} z)(x)
	+
	\sum_{f \subset \partial D_e}
	(\mathcal{K}^{\mathrm{flux}}_{e,f} z)(x),
	\quad x \in D_e,
	\label{eq:dgno_operator_local_cont}
\end{equation}
\begin{equation}
	(\mathcal{K}^{\mathrm{vol}}_e z)(x)
	=
	\int_{D_e} \kappa_\phi(x,y)\, z(y)\, dy,
%	\qquad x \in D_e.
	\label{eq:dgno_volume}
\end{equation}
\begin{equation}
	(\mathcal{K}^{\mathrm{flux}}_{e,f} z)(x)
	=
	\int_{f}
	\kappa_{\phi}(x,y)\,
	\widehat{\mathcal{F}}
	\!\left(
	z_e(y),\, z_{e'}(y)
	\right)\,
	dy.
%	\qquad x\in D_e,
	\label{eq:dgno_face}
\end{equation}

Here $(\mathcal{K}^{\mathrm{vol}}_{e} z)(x)$ denotes the element-local
kernel volume integral on the $D_e$, and
$(\mathcal{K}^{\mathrm{flux}}_{e,f} z)(x)$ represents the numerical flux
operator associated with the interface
$f \subset \partial D_e$ and its neighboring element.
The function $\widehat{\mathcal{F}}(\cdot,\cdot)$ is a numerical flux
that combines the interface information from both sides.
Specifically, $z_e(y)$ and $z_{e'}(y)$ denote the  evaluations of the latent
field on the interface $f$ taken from the interiors of the two adjacent
elements $D_e$ and $D_{e'}$, respectively.

%The numerical flux $\widehat{\mathcal{F}}(\cdot,\cdot)$ serves as the
%sole mechanism for cross-element interaction, mediating information
%exchange across element interfaces while preserving the element-wise
%representation.
%By combining the two traces, the flux operator enables controlled
%coupling between neighboring elements without imposing continuity
%constraints on the latent field.
%In classical DG formulations, numerical fluxes are designed analytically
%to ensure consistency and stability; in the proposed DG neural operator,
%$\widehat{\mathcal{F}}(\cdot,\cdot)$ is formulated at the operator level,
%allowing the interface coupling to be flexibly prescribed or learned
%from data.

\paragraph{Element-local Volume Integral Operator.}
Following the Galerkin-type attention formulation~\cite{wei2023super}, the integral operator is
parameterized through learned query, key, and value functions
$q(x)=W_q z(x)$, $k(x)=W_k z(x)$, and $v(x)=W_v z(x)\in\mathbb{R}^d$.
The element-local volume kernel integral operator Eq.~\eqref{eq:dgno_volume} then can be written in a component-wise form $(j = 1, \ldots, d_z)$ as
\begin{equation}
	\bigl((\mathcal{K}^{\mathrm{vol}}_e z)(x)\bigr)_j
	\approx
	\sum_{\ell=1}^{d_z}
	\left(
	\int_{D_e}
	k_\ell(y)\, v_j(y)\, dy
	\right)
	q_\ell(x),
%	x \in D_e,
	\label{eq:kernel_component_local}
\end{equation}
which expresses the output as a linear combination of the learned basis
functions $\{q_\ell(x)\}_{\ell=1}^{d_z}$, with coefficients given by
element-local inner products between key and value functions.
The Eq.~\eqref{eq:kernel_component_local} are approximated by a Monte-Carlo
quadrature over samples
$\{y_i^e\}_{i=1}^{n_e} \subset D_e$ as
\begin{equation}
	\int_{D_e} k_\ell(y)\, v_j(y)\, dy
	\approx
	\frac{1}{n_e}
	\sum_{i=1}^{n_e}
	k_\ell(y_i^e)\, v_j(y_i^e),
	\label{eq:mc_local}
\end{equation}
where $n_e=p^2$ for a $p\times p$ window.
%Let $Q_e, K_e, V_e\in\mathbb{R}^{h\times n_e\times d}$ denote the multi-head
%query, key, and value evaluations on element $D_e$, with $d=C/h$.
Let $Q_e, K_e, V_e \in \mathbb{R}^{h \times n_e \times d}$ denote the multi-head query, key, and value evaluations on element $D_e$ where $d = C/h$, and whose column vectors represent learned basis functions spanning subspaces of their respective latent representation Hilbert spaces.
The element-local volume integral operator is then obtained by applying the query representations
\begin{equation}
%	\hat{Z}_e
	(\mathcal{K}^{\mathrm{vol}}_{e} z)(x)
%	=
	\approx
	Q_e \mathbf{K}^{\mathrm{vol}}_e
	=
	\frac{1}{p^2}\,
	Q_e\, \tilde{K}_e^{\top} \tilde{V}_e,
	\label{eq:local_apply_combined}
\end{equation}
where $\tilde{K}_e = \mathrm{Ln}(K_e), \tilde{V}_e = \mathrm{Ln}(V_e)$ with
$\mathrm{Ln}(\cdot)$ denoting the layer normalization 
and $\mathbf{K}^{\mathrm{vol}}_e \in \mathbb{R}^{h \times d \times d}$
denotes the coefficient matrix of the basis functions for the element-local volume integral operator.

\paragraph{Interface-based Numerical Flux Operator.}
Analogously to the element-local volume integral operator, the flux operator uses the same Monte-Carlo Galerkin discretization but integrates over element interfaces instead of the entire element.
For a face $f \subset \partial D_e$ shared by neighboring elements
$D_e$ and $D_{e'}$, the Eq.~\eqref{eq:dgno_face} can be approximated as
\begin{equation}
\small
	(\mathcal{K}^{\mathrm{flux}}_{e,f} z)(x)
	\approx
	\frac{1}{n_f}
	\sum_{i=1}^{n_f}
	\kappa_{\phi}(x,y_i)\,
	\widehat{\mathcal{F}}
	\!\left(
	z_e(y_i),\, z_{e'}(y_i)
	\right),
%	\qquad x \in D_e,
	\label{eq:dgno_flux_mc}
\end{equation}
where $n_f$ is the number of sampled points sampled on the
interface $f$.
%and $z_e(y_i^f)$ and $z_{e'}(y_i^f)$ denote the traces of
%the latent field on the two sides of the interface.The trace of the latent field refers to its boundary value obtained by the element-wise representation to the interface from a given element.
Subsequently, the multi-head key and value evaluations on face $f$ are assembled into matrices
$K_{e,f}, V_{e,f} \in \mathbb{R}^{h \times n_f \times d}$. The
interface coefficient matrix is
\begin{equation}
	\mathbf{K}^{\partial,f}_{e}
	=
	\frac{1}{n_f}\,
	\tilde{K}_{e,f}^{\top} \tilde{V}_{e,f}
	\in \mathbb{R}^{h \times d \times d},
	\label{eq:dgno_boundary_operator}
\end{equation}
with $\tilde{K}_{e,f} = \mathrm{Ln}(K_{e,f})$ and
$\tilde{V}_{e,f} = \mathrm{Ln}(V_{e,f})$.
An analogous definition holds for
$\mathbf{K}^{\partial,f}_{e'}$ on the neighboring element $D_{e'}$ sharing the same face $f$.
Assuming that elements $D_e$ and $D_{e'}$ share the same basis functions on the shared face $f$, consistent with those used in the volume integral operator, the numerical flux operator can be written as
\begin{equation}
	(\mathcal{K}^{\mathrm{flux}}_{e,f} z)(x)
	\approx
	Q_e \mathbf{K}^{\mathrm{flux}}_{e,f}
	=
	Q_e \widehat{\mathcal{F}}
	\!\left(
	\mathbf{K}^{\partial,f}_{e},
	\mathbf{K}^{\partial,f}_{e'}
	\right),
%	\in \mathbb{R}^{h \times d \times d}.
	\label{eq:dgno_flux_operator}
\end{equation}
where $\mathbf{K}^{\mathrm{flux}}_{e,f}$ denotes the coefficient matrix of the numerical flux on element $D_e$.
This implies that each local element not only captures its own internal information but also exchanges information with neighboring elements through numerical fluxes, thereby enabling global coupling.
%The contribution of face $f$ to the element-local update is obtained by
%applying the query projection,
%\begin{equation}
%	\hat{Z}^{\,\mathrm{flux}}_{e,f}
%	=
%	Q_e\, \mathbf{K}^{\mathrm{flux}}_{e,f}.
%	\label{eq:dgno_flux_apply}
%\end{equation}

\paragraph{P0DG-based Numerical Flux Operator.}
Beyond the interface-based flux construction described above, the
discontinuous Galerkin neural operator also admits a Zero-order DG
(P0DG) formulation, in which the numerical flux is derived directly
from element-local volume integral operators.
Since the latent field is represented as piecewise constant on each element and exhibits no spatial variation within the element, the element-local volume operator fully characterizes the element’s contribution, enabling the numerical flux across interfaces to be computed directly from the neighboring cell-wise representations without the need for any additional interface integral operations.
Consequently, the numerical flux operator in the P0DG case is given by
\begin{equation}
	\mathbf{K}^{\mathrm{flux}}_{e,f}
	=
	\widehat{\mathcal{F}}
	\!\left(
	\mathbf{K}^{\mathrm{vol}}_{e},
	\mathbf{K}^{\mathrm{vol}}_{e'}
	\right)
	\in \mathbb{R}^{h \times d \times d},
	\label{eq:dgno_flux_operator_P0}
\end{equation}
where $\mathbf{K}^{\mathrm{vol}}_{e}$ and
$\mathbf{K}^{\mathrm{vol}}_{e'}$ denote the coefficient matrices of the element-local volume
operators on the two neighboring elements.
This P0DG formulation can be interpreted as a lowest-order approximation of the interface-based flux.

\paragraph{Discrete DG Neural Operator Assembly.}
Combining the element-local volume operator with all interface-based flux
contributions, the latent update in Eq.~\eqref{eq:dgno_operator_local_cont} is given by
\begin{equation}
\small
%	\hat{Z}_e
	(\mathcal{K}^{\mathrm{DG}}_{e} z)(x)
%	=
	\approx
	Q_e
	\left(
	\mathbf{K}^{\mathrm{vol}}_{e}
	+
	\sum_{f \subset \partial D_e}
	\mathbf{K}^{\mathrm{flux}}_{e,f}
	\right)
	=
	Q_e\, \mathbf{K}^{\mathrm{DG}}_{e},
	\label{eq:dgno_element_update}
\end{equation}
where
\(
\mathbf{K}^{\mathrm{DG}}_{e}
\in \mathbb{R}^{h \times d \times d}
\)
denotes the discrete DG coefficient matrix on element $D_e$. 
Finally, by imposing boundary conditions and numerical fluxes at the operator level, the contributions from all elements are assembled to produce the global DG representation of the neural operator. 
Four representative numerical fluxes, namely central, upwind, jump, and average-jump, are considered under three types of boundary conditions, including Neumann, Dirichlet, and periodic; implementation details are provided in Appendix~\ref{app:Fluxes} and Appendix~\ref{app:boundary_conditions}.
In this paper, DGNO with an interface-based flux is referred to as \textbf{DGNO-Face}, while DGNO with a P0DG-based flux is denoted as \textbf{DGNO-Cell}.

\textbf{Network details.}
The overall network architecture is illustrated in Fig.~\ref{fig:DGNO_arch}.
DGNO first lifts the blurred input image $x$ into a multi-scale feature space, producing feature maps with channel dimensions $d_e = {48, 96, 192, 384}$ across four scales.
The encoder captures shared basis functions from the training distribution, while the proposed discontinuous Galerkin-type attention layers further enable instance-specific basis refinement.
We adopt the multi-head attention mechanism~\cite{vaswani2017attention} by partitioning the queries, keys, and values into $n_{\text{heads}}$ independent heads, each with dimensionality $d_z / n_{\text{heads}}$.
In our implementation, we set $d_z = {48, 96, 192}$ and $n_{\text{heads}} = 16$, resulting in 3-, 6-, and 12-dimensional features per head, respectively.
The highest-scale features (384 channels) are upsampled and fused with the 192-channel features before being passed to the integral operator.
We employ only two iterations ($T = 2$) of the kernel integral operator, which already outperforms prior methods while maintaining high computational efficiency\vspace{-0.3em}.

\begin{table*}
	\renewcommand\arraystretch{0.8}
	\centering
	\caption{Comparisons with other Single Image Defocus Deblurring methods on BBBC006$_{w1}$~\cite{ljosa2012annotated}, BBBC006$_{w2}$~\cite{ljosa2012annotated} and 3DHistech~\cite{geng2022cervical}.}
	\label{resultmicro}
	\vspace{-0.5em}
	% \fontsize{8.8pt}{6.5pt}\selectfont
	\scalebox{0.9}{
		\begin{tabular}{lc@{}lc@{}lc@{}lc@{}lc@{}}
			\toprule
			\multirow{2}{*}{Method} & \multicolumn{3}{c}{BBBC006$_{w1}$} & \multicolumn{3}{c}{BBBC006$_{w2}$} & \multicolumn{3}{c}{3DHistech} \\
			%			\cmidrule{2-10}
			\cmidrule(lr){2-4} \cmidrule(lr){5-7} \cmidrule(lr){8-10}
			& \multicolumn{1}{c}{PSNR$\uparrow$} & \multicolumn{1}{c}{SSIM$\uparrow$} & \multicolumn{1}{c}{LPIPS$\downarrow$} & \multicolumn{1}{c}{PSNR$\uparrow$} & \multicolumn{1}{c}{SSIM$\uparrow$} & \multicolumn{1}{c}{LPIPS$\downarrow$} & \multicolumn{1}{c}{PSNR$\uparrow$} & \multicolumn{1}{c}{SSIM$\uparrow$} & \multicolumn{1}{c}{LPIPS$\downarrow$} \\
			\midrule
			GKMNet~\cite{quan2021gaussian} & \multicolumn{1}{c}{34.42} & \multicolumn{1}{c}{0.941} & \multicolumn{1}{c}{0.132} & \multicolumn{1}{c}{26.87} & \multicolumn{1}{c}{0.785} & \multicolumn{1}{c}{0.396} & \multicolumn{1}{c}{33.42} & \multicolumn{1}{c}{0.852} & \multicolumn{1}{c}{0.130} \\
			MIMO-Unet~\cite{cho2021rethinking} & \multicolumn{1}{c}{35.15} & \multicolumn{1}{c}{0.948} & \multicolumn{1}{c}{0.117} & \multicolumn{1}{c}{29.70} & \multicolumn{1}{c}{0.828} & \multicolumn{1}{c}{0.354} & \multicolumn{1}{c}{32.40} & \multicolumn{1}{c}{0.837} & \multicolumn{1}{c}{0.169} \\
			NAFNet~\cite{chen2022simple} & \multicolumn{1}{c}{34.98} & \multicolumn{1}{c}{0.947} & \multicolumn{1}{c}{0.111} & \multicolumn{1}{c}{29.44} & \multicolumn{1}{c}{0.825} & \multicolumn{1}{c}{0.341} & \multicolumn{1}{c}{33.23} & \multicolumn{1}{c}{0.889} & \multicolumn{1}{c}{0.132} \\
			SwinIR~\cite{liu_swin_2021} & \multicolumn{1}{c}{34.75} & \multicolumn{1}{c}{0.943} & \multicolumn{1}{c}{0.123} & \multicolumn{1}{c}{30.02} & \multicolumn{1}{c}{0.829} & \multicolumn{1}{c}{0.351} & \multicolumn{1}{c}{32.57} & \multicolumn{1}{c}{0.841} & \multicolumn{1}{c}{0.136}\\
			
			MambaIRv2~\cite{guo2025mambairv2} & \multicolumn{1}{c}{34.35} & \multicolumn{1}{c}{0.941} & \multicolumn{1}{c}{0.113} & \multicolumn{1}{c}{30.98} & \multicolumn{1}{c}{0.823} & \multicolumn{1}{c}{0.361} & \multicolumn{1}{c}{33.48} & \multicolumn{1}{c}{0.881} & \multicolumn{1}{c}{0.106}  \\
			%			P-former~\cite{cai2025pathology} & \multicolumn{1}{c}{-} & \multicolumn{1}{c}{-} & \multicolumn{1}{c}{-} & \multicolumn{1}{c}{-} & \multicolumn{1}{c}{-} & \multicolumn{1}{c}{-} & \multicolumn{1}{c}{\underbar{33.68}} & \multicolumn{1}{c}{\textbf{0.903}} & \multicolumn{1}{c}{{0.102}} \\

			Restormer~\cite{zamir2022restormer} & \multicolumn{1}{c}{35.35} & \multicolumn{1}{c}{{0.950}} & \multicolumn{1}{c}{\textbf{0.103}} & \multicolumn{1}{c}{{31.70}} & \multicolumn{1}{c}{{0.842}} & \multicolumn{1}{c}{{0.331}} & \multicolumn{1}{c}{33.46} & \multicolumn{1}{c}{0.880} & \multicolumn{1}{c}{0.125} \\
			%			Loformer~\cite{mao2024loformer} & \multicolumn{1}{c}{-} & \multicolumn{1}{c}{-} & \multicolumn{1}{c}{-} & \multicolumn{1}{c}{-} & \multicolumn{1}{c}{-} & \multicolumn{1}{c}{-} & \multicolumn{1}{c}{-} & \multicolumn{1}{c}{-} & \multicolumn{1}{c}{-} & \multicolumn{1}{c}{-} & \multicolumn{1}{c}{-} & \multicolumn{1}{c}{-} \\
			
			MPT+EFCR~\cite{zhang2024unified} & \multicolumn{1}{c}{{35.44}} & \multicolumn{1}{c}{0.947} & \multicolumn{1}{c}{0.114} & \multicolumn{1}{c}{30.48} & \multicolumn{1}{c}{0.830} & \multicolumn{1}{c}{0.348} & \multicolumn{1}{c}{33.58} & \multicolumn{1}{c}{0.887} & \multicolumn{1}{c}{0.119} \\

			\cmidrule{1-10}
			DGNO-Face & \multicolumn{1}{c}{\underbar{37.09}} & \multicolumn{1}{c}{\underbar{0.958}} & \multicolumn{1}{c}{\underbar{0.104}} & \multicolumn{1}{c}{\textbf{32.66}} & \multicolumn{1}{c}{\underbar{0.847}} & \multicolumn{1}{c}{\underbar{0.323}} & \multicolumn{1}{c}{\textbf{34.02}} & \multicolumn{1}{c}{\textbf{0.890}} & \multicolumn{1}{c}{\underbar{0.095}} \\
			
			DGNO-Cell & \multicolumn{1}{c}{\textbf{37.22}} & \multicolumn{1}{c}{\textbf{0.959}} & \multicolumn{1}{c}{\textbf{0.103}} & \multicolumn{1}{c}{\underbar{32.54}} & \multicolumn{1}{c}{\textbf{0.848}} & \multicolumn{1}{c}{\textbf{0.322}} & \multicolumn{1}{c}{\underbar{34.00}} & \multicolumn{1}{c}{\textbf{0.890}} & \multicolumn{1}{c}{\textbf{0.093}} \\
			
			\bottomrule
		\end{tabular}
	}
	\vspace{-0.8em}
	%	\caption{Comparisons with other Single Image Defocus Deblurring methods on BBBC006$_{w1}$~\cite{ljosa2012annotated}, BBBC006$_{w2}$~\cite{ljosa2012annotated} and 3DHistech~\cite{geng2022cervical}}
	%	\label{resultmicro}
\end{table*}

\begin{figure*}[!htb]
	\centering
	\vspace{-0.2em}
	\begin{center}
		\includegraphics[width=0.85\textwidth]{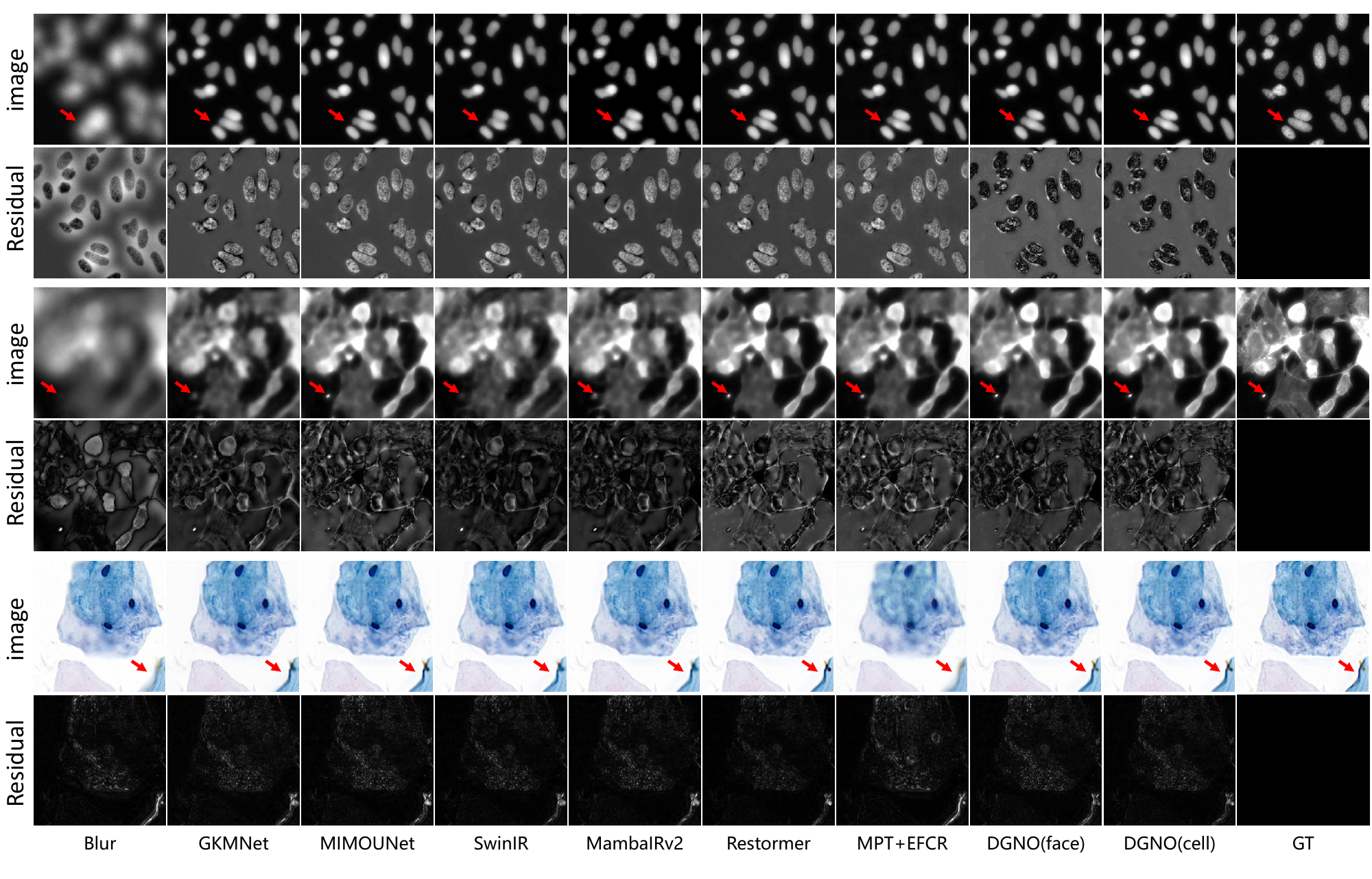}
		\vspace{-0.7em}
		\caption{
			Visual comparison of single image defocus deblur approaches on BBBC006$_{w1}$ and BBBC006$_{w2}$.}
		\label{fig:DGNO_Visual}
	\end{center}
	%	\vspace{-20pt}
	\vspace{-2.3em}
\end{figure*}

\section{Experiments}
To demonstrate the effectiveness of our model, following MPT~\cite{zhang2024unified}, we evaluate DGNO on three microscopic defocus blur datasets: BBBC006$_{w1}$~\cite{ljosa2012annotated}, BBBC006$_{w2}$~\cite{ljosa2012annotated}, 3DHistech~\cite{geng2022cervical}. We provide more details of the used datasets, training settings, and additional visual results in Appendix.  
%In tables, the best result is highlighted in \textbf{bold}.

\textbf{Implementation details.} The DGNO model adopts a multi-scale neural-operator architecture with four levels of lifting and three output scales. The number of encoder modules \cite{guo2025mambairv2} from first to fourth level is $[2, 4, 6, 2]$ with channel dimensions $[48, 96, 192, 384]$. DGNO operates on non-overlapping local elements of size $8 \times 8$. The model was trained on NVIDIA RTX4090 GPUs (48 GB) using AdamW optimizer ($\beta_1=0.9, \beta_2=0.999$, weight decay=$1\times10^{-4}$) with an initial learning rate of $3\times10^{-4}$ (cosine decay to $1\times10^{-6}$). The batch size is 8 with training patches in the size of $256\times256$ image patches with data augmentation including horizontal flipping and vertical flipping.

\subsection{Comparison with State-of-the-Art}
We compare the proposed DGNO with representative state-of-the-art
single-image defocus deblurring methods on BBBC006~\cite{ljosa2012annotated} and 3DHistech~\cite{geng2022cervical}, as summarized in Table~\ref{resultmicro}.
On BBBC006$_{w1}$, DGNO achieves the best performance, reaching 37.22/37.09 dB with 50 FLOPs and 12.79M parameters, exceeding Restormer and MPT+EFCR by up to 1.87 dB.
On the more challenging BBBC006$_{w2}$ dataset, DGNO further outperforms MPT and Restormer by 2.18 dB and 0.96 dB, respectively.
On the 3DHistech dataset, both DGNO-Face and DGNO-Cell yield similar PSNR results, and both outperform MPT+EFCR by 0.44 dB.
Note that DGNO-Cell outperforms DGNO-Face on BBBC006${w1}$, whereas the opposite trend is observed on the more challenging BBBC006${w2}$ dataset.
This observation suggests that DGNO-Cell, owing to its piecewise-constant approximation, is better suited to relatively simpler imaging scenarios, while DGNO-Face is more effective in handling more complex structural patterns. More discussions regarding stability and generalizability on DGNO-Cell and DGNO-Face are discussed in Sec.~\ref{sec:AD}.

\subsection{Analysis and Discussion}
\label{sec:AD}
This section analyzes the proposed DGNO framework by comparing discontinuous
Galerkin operators with global Galerkin operators, examining numerical flux and
boundary condition designs, discussing generalization to real-world defocus and evaluating generalization across lifting
modules, defocus levels, and downstream tasks.

\paragraph{From Global Galerkin Operators to Discontinuous Galerkin Operators.}
Table~\ref{compared_global} presents an ablation study comparing the proposed DGNO with a global Galerkin (GG) operator (SRNO) and its local windowed variant on BBBC006$_{w1}$. Although GG was initially developed for image super-resolution \cite{wei2023super}, aiming at learning global smoothness and arbitrary upsampling, we are the first to explore its performance in pathological image defocus deblurring from the physical imaging process perspective.
As a strong baseline, the GG operator achieves a PSNR of 36.71 dB, outperforming the Transformer-based MPT (35.44 dB), thereby highlighting the advantage of neural operator–based modeling for defocus deblurring.
Building upon this baseline, introducing a window-based local Galerkin (LG) operator (SRNO+Win) yields a PSNR gain of 0.14 dB, indicating the benefit of enforcing locality consistent with the local integral nature of defocus blur.
Going beyond windowed locality, DGNO with interface-based numerical flux further improves the PSNR to 37.07 dB without increasing the computational complexity.
Finally, adopting the P0DG-based flux achieves the best performance across all metrics, demonstrating that DG operators provide a more faithful operator-level modeling of defocus blur.

\begin{table}[t]
	\caption{Comparison with the global Galerkin operator on BBBC006$_{w1}$~\cite{ljosa2012annotated}.}
	\label{compared_global}
	\vspace{-0.8em}
	\fontsize{8.8pt}{6.5pt}\selectfont
	\setlength{\tabcolsep}{1pt}
	\renewcommand\arraystretch{0.6}
	\begin{center}
		\begin{small}
			\begin{sc}
				\begin{tabular}{lccc|cc}
					\toprule
					&\multicolumn{3}{c}{{BBBC006$_{w1}$}} & {Params} & {FLOPs}  \\
					\cmidrule(lr){2-4} \cmidrule(lr){5-6} 
					{Method}           & PSNR         & SSIM      & LPIPS & (MB) & (G)  \\
					\midrule
					SRNO(GG)    & 36.71  & 0.957 & 0.104 & 12.79 & 50.12  \\
					SRNO+Win(LG) & 36.85  & 0.957 & 0.104 & 12.79 & 50.12 \\
					\midrule
					DGNO-Face      & \underbar{37.07} & \underbar{0.958} & \underbar{0.104} & 12.79 & 50.15  \\
					DGNO-Cell      & \textbf{37.21} & \textbf{0.959} & \textbf{0.103} & 12.79 & 50.12  \\
					\bottomrule
				\end{tabular}
			\end{sc}
		\end{small}
	\end{center}
	% \vskip -0.1in
	\vspace{-2.3em}
\end{table}

\begin{figure}[!htb]
	\centering
	\vspace{-1em}
	\begin{center}
		\includegraphics[width=0.48\textwidth]{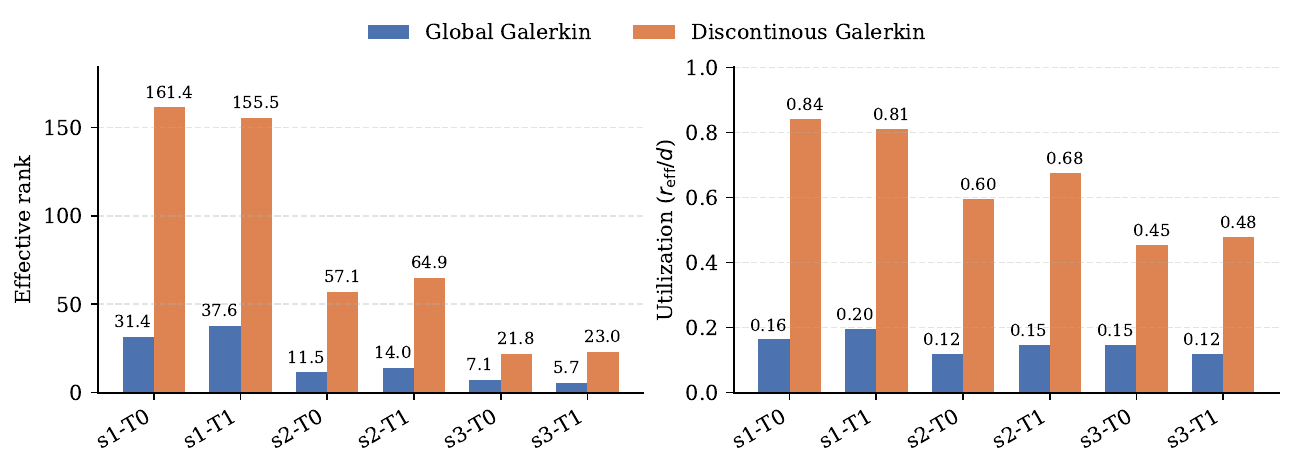}
		\vspace{-2em}
		\caption{Effective-rank analysis of latent representations across scales and iteration steps ($T=0,1$). (a) the absolute effective rank of the latent matrix $z$, (b) the effective-rank utilization $r_{\mathrm{eff}}/d$
		}
		\label{fig:rank}
	\end{center}
	\vspace{-1.4em}
\end{figure}

To validate the discontinuous pattern learning ability of DGNO, as shown in Fig.~\ref{fig:rank},  we evaluate the representation capacity of GG and DG operator via the effective rank of the latent matrix $z$ and its utilization. \textbf{Note that the higher the rank, the more {discontinuous patterns} the model can learn.}
Across all scales ($s1$: $H/4 \times W/4$, 192-dim; $s2$: $H/2 \times W/2$, 96-dim; $s3$:
$H \times W$, 48-dim) and iteration steps ($T=0,1$), DGNO-Face consistently achieves higher
effective rank and utilization than the GG operator.
These results demonstrate that DGNO enriches basis
diversity and prevent rank collapse during iterative operator refinement.

\begin{figure}[t]
	\centering
	\begin{center}
		\includegraphics[width=0.48\textwidth]{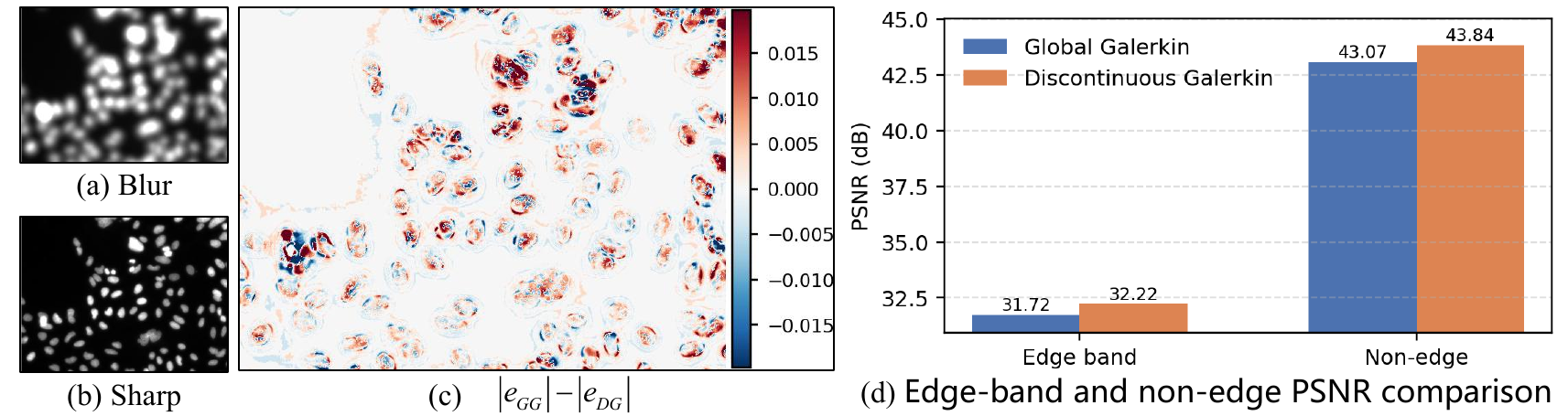}
		\caption{Residual visualization and edge-aware quantitative analysis comparing
			GG and DG operators.
			(a) Blur image
			(b) Sharp image.
			(c) Absolute error difference
			$\lvert e_{\mathrm{GG}} \rvert - \lvert e_{\mathrm{DG}} \rvert$.
			(d) PSNR comparison on edge-band and non-edge regions.}
		\label{fig:vis_boundary}
	\end{center}
	\vspace{-15pt}
\end{figure}

\begin{table}[t]
\renewcommand\arraystretch{0.9}
	\caption{Ablation on Different Flux and Boundary Condition. Neumann Boundary (NB), Dirichlet Boundary (DB) and Periodic Boundary (PB).}
    \vspace{-1em}
	\label{flux_boundary}
    \fontsize{8.8pt}{6.5pt}\selectfont
	\setlength{\tabcolsep}{3pt}
	\begin{center}
		\begin{small}
			\begin{sc}
            %\resizebox{0.95\linewidth}{!}{
				\begin{tabular}{l|ccc|ccc}
					\toprule
					&\multicolumn{3}{c}{DGNO-Face} & \multicolumn{3}{c}{DGNO-Cell} \\
					Flux & NB & DB & PB & NB & DB & PB  \\
					\midrule
					Central      & 36.98 & 36.99 & 36.86 & 36.91   & 37.06  & 36.88   \\
					Jump         & 37.02 & 37.06 & 37.06 & \textbf{37.22}   & 37.16  & 36.97   \\
					Avg-jump     & 37.02 & \textbf{37.09} & 36.99 & 37.02   & 36.85  & 37.01    \\
					Upwind       & 36.95 & 37.03 & 36.88 & 37.08   & 37.03  & 36.92  \\
					\bottomrule
				\end{tabular}
			\end{sc}
		\end{small}
	\end{center}
	% \vskip -0.1in
    \vspace{-2em}
\end{table}

To further analyze the boundary-aware behavior of the DG operator, Fig.~\ref{fig:vis_boundary} presents a qualitative and quantitative comparison with the GG operator on the BBBC006$_{w1}$ dataset.
The error difference map in Fig.~\ref{fig:vis_boundary}(c), defined as $\lvert e_{\mathrm{GG}} \rvert - \lvert e_{\mathrm{DG}} \rvert$, where $\lvert e_{\mathrm{GG}} \rvert$ and $\lvert e_{\mathrm{DG}} \rvert$ denote the absolute reconstruction errors of the GG and DG operators, respectively, shows that DG operator consistently reduces reconstruction errors near object boundaries and fine cellular structures, while maintaining comparable accuracy in homogeneous regions.
The region-wise PSNR analysis in Fig.~\ref{fig:vis_boundary}(d) further confirms this trend.
By separately evaluating edge-band and interior regions, DG operator achieves a PSNR improvement of $+0.49~\mathrm{dB}$ on boundary regions and also a gain of $+0.77~\mathrm{dB}$ within non-edge regions.

\begin{figure}[t]
	\centering
	\begin{center}
		\includegraphics[width=0.45\textwidth]{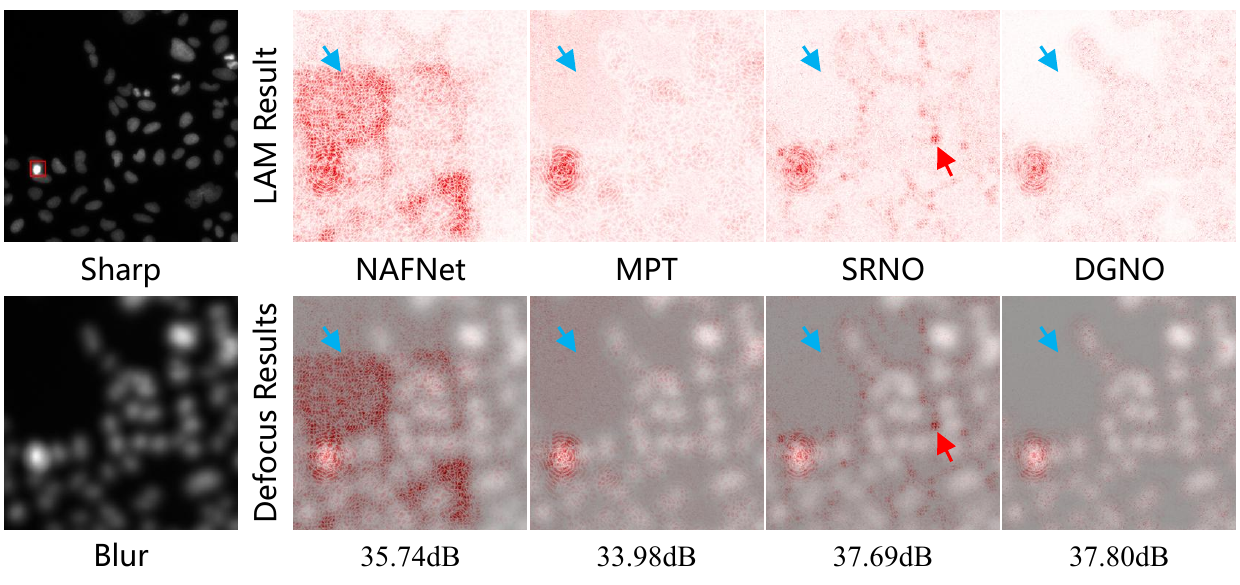}
		\caption{LAM-based interpretability and defocus deblurring results on BBBC006.}
		\label{fig:e4}
	\end{center}
	\vspace{-2em}
\end{figure}

To analyze the interpretability and qualitative behavior of different defocus deblurring models, Fig.~\ref{fig:e4} presents a visual comparison of local attribution maps (LAM) \cite{gu2021interpreting} on the BBBC006$_{w1}$ dataset.
The LAM provides spatial interpretability by revealing how each model distributes its responses across image regions.
Compared with CNN- and Transformer-based baselines such as NAFNet and MPT, as well as the global Galerkin operator SRNO, the proposed DGNO exhibits more localized and boundary-aligned activations.
Specifically, NAFNet exhibits activations that are distributed within a large region, which may correlate with its convolutional inductive bias, and is inconsistent with the physical image formation process. MPT shows sparse yet spatially diffuse responses, such as the activations in the upper-left region (see the blue arrow), which may correlate with the global learning ability of self-attention. 
In contrast, globally coupled models SRNO exhibit spatially diffuse activations indicative of excessive long-range mixing and boundary interference, as highlighted by the red arrows in Fig.~\ref{fig:e4}.
Correspondingly, the more localized and boundary-aligned activations produced by DGNO (highlighted by the blue arrows in Fig.~\ref{fig:e4}) lead to sharper reconstruction of cellular boundaries and finer internal structures, resulting in the highest PSNR among all methods.
These results show the discontinuous Galerkin formulation effectively enables controlled cross-region interactions via interface fluxes while preserving element-wise locality, leading to superior boundary fidelity and overall reconstruction quality.

\begin{figure}[!htb]
	\centering
	\begin{center}
		\includegraphics[width=0.35\textwidth]{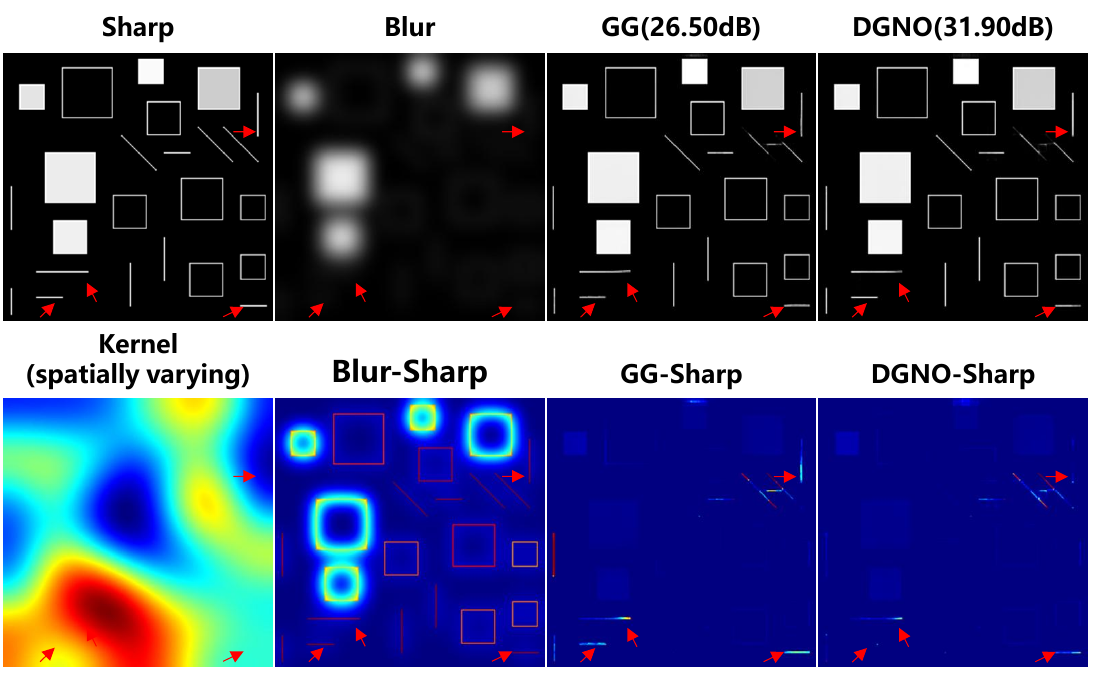}
		\caption{Visual comparison on a controlled synthetic example with spatially varying Gaussian blur.}
		\label{fig:e5}
	\end{center}
	\vspace{-1.5em}
\end{figure}

\paragraph{Controlled Synthetic Experiments under Spatially Varying Blur.}
To more directly validate the advantage of DGNO under spatially varying blur, we conduct two controlled synthetic experiments. First, on 50 randomly generated synthetic images, we constructed sharp patterns composed of filled squares, hollow boxes, and thin lines, and applied known spatially varying Gaussian blur, where the blur strength at each pixel was controlled by a predefined sigma map. This setting isolates restoration performance under spatially heterogeneous blur with fully known degradation kernels, without interference from real-image distribution bias. DGNO consistently outperforms the global Galerkin baseline under different blur ranges: when the sigma range is 8--10, the average PSNR of GG, DGNO-Face, and DGNO-Cell is 34.69, 38.96, and 37.91 dB, respectively; when the range is expanded to 0.6--13, the corresponding values are 44.24, 46.23, and 46.32 dB. As shown in Fig.~\ref{fig:e5}, GG leaves more residual blur and loses local structures around thin lines, hollow-box boundaries, and other high-frequency regions, whereas DGNO restores sharper edges and exhibits weaker residual errors. Second, on BBBC006$_{w1}$ (153 images), we further synthesized spatially varying blur using known Gaussian kernels with random spatial variation and a sigma range of 0.6--13. Under this setting, DG improves PSNR over GG by 1.94 dB (25.81 to 27.75 dB), and still achieves a 1.77 dB gain on boundary regions, indicating stronger modeling ability in spatial transition areas.

\begin{figure}[!htb]
	\centering
	\vspace{-0.5em}
	\begin{center}
		\includegraphics[width=0.45\textwidth]{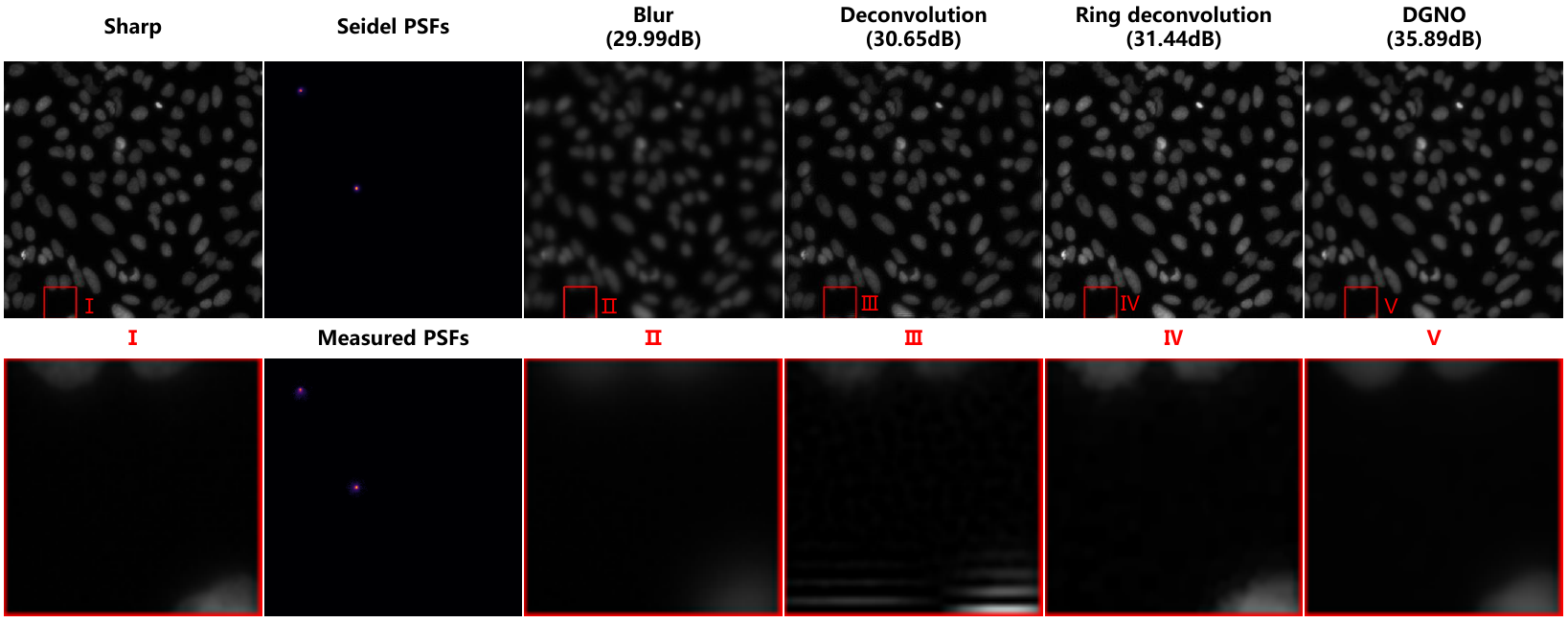}
		\caption{Visual comparison with non-blind spatially varying deconvolution on BBBC006$_{w1}$ under Seidel-coefficient-based blur.}
		\label{fig:e6}
	\end{center}
	\vspace{-2.3em}
\end{figure}

\paragraph{Comparison with Non-blind Spatially Varying Deconvolution.}
To evaluate performance under more realistic spatially varying blur, we compare our method on BBBC006$_{w1}$ with ring deconvolution~\cite{kohli2025ring}, a non-blind spatially varying deconvolution method. The spatially varying PSF is generated from Seidel coefficients estimated from a calibration image using rdmpy, and blurred observations are synthesized via the spatially varying forward model. Table~\ref{non_blind} compares our method with the blurred input, a non-blind deconvolution baseline using only a single center PSF, and Ring deconvolution. Our method achieves 36.51 dB / 0.937 SSIM, outperforming both the blurred input (31.07 dB / 0.873) and the single-PSF baseline (32.87 dB / 0.920). Ring deconvolution performs best (40.15 dB / 0.966), serving as a non-blind upper bound with access to the ground-truth spatially varying PSF. This result validates the effectiveness of our method in handling spatially varying blur without explicit PSF access.
As shown in Fig.~\ref{fig:e6}, the blurred image exhibits severe spatially varying degradation, while single-PSF deconvolution only provides limited recovery and introduces noticeable artifacts in the zoomed region. Ring deconvolution restores sharper structures with access to the true spatially varying PSF, whereas our blind DGNO recovers nuclei boundaries and fine structures.

\begin{table}[t]
	\caption{Quantitative comparison on BBBC006$_{w1}$~\cite{ljosa2012annotated} with Seidel-coefficient-based spatially varying blur.}
	\label{non_blind}
	\centering\small
%	\vspace{-0.3em}
	\renewcommand\arraystretch{0.9}
	\resizebox{1\linewidth}{!}{
		\begin{tabular}{l|cc|cc}
			\toprule
			Model           & Blind         & Uses true spatially varying PSF      & PSNR & SSIM  \\
			\midrule
			Blur          & - & - & 31.07 & 0.873 \\
			Deconvolution & $\times$ & $\times$(single center PSF only) & 32.87 & 0.920  \\
			Ring Deconvolution    & $\times$ & \checkmark & 40.15 & 0.966  \\
			\midrule
			Ours          & \checkmark & $\times$ & 36.51 & 0.937 \\
			\bottomrule
	\end{tabular}}
	\vspace{-2em}
\end{table}

\paragraph{Numerical Flux and Boundary Condition Analysis.}
Table~\ref{flux_boundary} presents an ablation study of numerical fluxes and
boundary conditions for DGNO-Face and DGNO-Cell on BBBC006$_{w1}$.
For DGNO-Face, the average-jump flux under Dirichlet boundary
conditions achieves the highest PSNR, whereas for DGNO-Cell, the jump
flux under Neumann boundary conditions yields the best performance.
Across different flux and boundary settings, DGNO-Face exhibits
more stable performance, while DGNO-Cell shows larger variability and
higher sensitivity to boundary conditions.
Table~\ref{Penalty} reports an ablation on the penalty coefficient for
both interface-based and cell-based numerical fluxes under their optimal
settings.
A learnable penalty consistently achieves the best PSNR, demonstrating the
effectiveness of adaptively controlling operator-level flux strength.

\textbf{Generalization to Real-World Defocus.}
%We also evaluate our method for real-world deblur on DPDD~\cite{abuolaim2020defocus}. Our method achieves the best performance in terms of SSIM and PSNR. It shows that our model is universally applicable to different types of images.
We further evaluate DGNO on real-world DPDD~\cite{abuolaim2020defocus}. 
As shown in Table~\ref{tab:dpdd}, DGNO-Face achieves the best PSNR and competitive perceptual quality among the compared methods, with 26.42 dB PSNR and 0.173 LPIPS.
The result demonstrates that the proposed operator-level formulation is not limited to pathology images and can generalize to real-world defocus blur in natural scenes. 
%As shown in Table~\ref{tab:dpdd}, DGNO-Face achieves the best PSNR and LPIPS, indicating that our operator-level formulation generalizes beyond pathology images to natural-scene defocus deblurring.

\begin{table}[!htb]
	\vspace{-0.6em}
	\caption{Quantitative real-world deblur evaluation on DPDD~\cite{abuolaim2020defocus}.}
	\label{tab:dpdd}
    \centering\small
    \vspace{-0.8em}
            \renewcommand\arraystretch{0.9}
            \resizebox{0.9\linewidth}{!}{
				\begin{tabular}{l|cccr}
					\toprule
					Model           & PSNR         & SSIM      & LPIPS  \\
					\midrule
					GKMNet~\cite{chen2022simple}         & 25.47 & 0.789 & 0.219 \\
					Restormer~\cite{zamir2022restormer}      & 25.98 & 0.811 & 0.178  \\
					MambaIR~\cite{guo2024mambair} & 26.11 & 0.809 & 0.202  \\
					MPT~\cite{zhang2024unified}       	   & 26.21 & 0.826 & 0.175  \\
					FSNet~\cite{cui2023image} & 26.22 & 0.811 & 0.207 \\
					ConvIR \cite{cui2024revitalizing}	   & 26.36 & 0.820 & 0.174 \\
					\midrule
					DGNO-Face & 26.42 & 0.814 & 0.173 \\
					\bottomrule
				\end{tabular}}
    \vspace{-1em}
\end{table}

\textbf{Generality across Lifting Modules for Neural Operators.}
Table~\ref{encoder} evaluates the generality of DGNO by instantiating the lifting operator with different backbone encoders.
Across all lifting module choices, DGNO consistently yields performance improvements.

\begin{table}[!htb]
\renewcommand\arraystretch{0.8}
\centering
\vspace{-0.8em}
	\caption{Ablation with Different Lifting Modules.}
	\label{encoder}
	\vspace{-0.8em}
    %\fontsize{8.8pt}{6.5pt}\selectfont
	%\setlength{\tabcolsep}{1pt}
    \small
    \resizebox{0.9\linewidth}{!}{
				\begin{tabular}{l|cccr}
					\toprule
					Model           & PSNR         & SSIM      & LPIPS  \\
					\midrule
					NAFNet~\cite{chen2022simple}         & 34.98 & 0.947 & 0.111 \\
					NAFNet+DGNO    & \textbf{35.72} & \textbf{0.953} & \textbf{0.989} \\
					Restormer~\cite{zamir2022restormer}      & 35.35 & 0.950 & 0.103  \\
					Restormer+DGNO & \textbf{35.53} & \textbf{0.952} & \textbf{0.102}  \\
					MPT~\cite{zhang2024unified}       	   & 35.44 & 0.947 & 0.114  \\
					MPT+DGNO 	   & \textbf{35.80} & \textbf{0.948} & \textbf{0.110} \\
					\bottomrule
				\end{tabular}}
	% \vskip -0.1in
    \vspace{-1em}
\end{table}

\textbf{Performance under Different Defocus Levels.}
Table~\ref{gene_w1} reports a quantitative comparison of different algorithms under decreasing defocus blur levels ($z=01, 05, 13$) on BBBC006$_{w1}$, where $z=16$ corresponds to the in-focus plane.
As the defocus severity decreases, all methods benefit from reduced blur, leading to consistent improvements in PSNR.
DGNO-Face consistently achieves the highest PSNR across all defocus levels, demonstrating strong robustness to varying blur conditions.
DGNO-Cell delivers competitive performance and outperforms most baselines; however, under stronger defocus, its effectiveness is constrained by the P0DG, which lacks explicit interface modeling and loses fine boundary information compared with DGNO-Face.
Overall, these results indicate that by learning the inverse integral kernel through a neural operator, DGNO enables interpretable defocus deblurring with strong generalization and stability.The corresponding results on BBBC006$_{w2}$ are provided in Table~\ref{gene_w2} in the appendix.

%\begin{table}[t]
%	\caption{Performance comparison of different algorithms under decreasing levels of defocus blur ($z=01, 05, 13$) on BBBC006$_{w1}$.}
%	\label{gene}
%	\setlength{\tabcolsep}{1pt}
%	\begin{center}
%		\begin{small}
%			\begin{sc}
%				\resizebox{\linewidth}{!}{
%					\begin{tabular}{l|cc|cc|cc}
%						\toprule
%						&\multicolumn{2}{c}{z=01} & \multicolumn{2}{c}{z=05} & \multicolumn{2}{c}{z=13} \\
%						Method& PSNR & SSIM & PSNR & SSIM & PSNR & SSIM \\
%						\midrule
%						GKMNet  & 31.50 & 0.911 & 33.43 & 0.933 & 32.78 & 0.939\\
%						NAFNet  & 32.37 & 0.928 & 34.43 & 0.943 & 36.33 & 0.947\\
%						MambaIRv2 &  30.63 & 0.900 & 33.25 & 0.932 & 36.29 & 0.949 \\
%						SwinIR~\cite{liu_swin_2021} & 30.32 &0.898  & 32.66 &0.925 & 34.21 & 0.912  \\
%						Restormer~\cite{zamir2022restormer} & 32.18 & 0.917 & 33.79 & 0.937 & 36.33 &0.947  \\
%						MPT~\cite{zhang2024unified} & 32.87 & 0.923 & 34.72 & 0.942  & 35.56 &0.937  \\
%						DGNO(cell) & 32.41 & 0.919 & 34.87 & 0.945 & 36.25 &0.939  \\
%						DGNO(face) & \textbf{34.30} & \textbf{0.940} & \textbf{36.32} & \textbf{0.953} & \textbf{36.99} & \textbf{0.947}  \\
%						\bottomrule
%				\end{tabular}}
%			\end{sc}
%		\end{small}
%	\end{center}
%	\vskip -0.1in
%\end{table}

\begin{table}[t]
\renewcommand\arraystretch{0.95}
\small
	\caption{Generalization performance of different algorithms under decreasing levels of defocus blur ($z=01, 05, 13$) on BBBC006$_{w1}$~\cite{ljosa2012annotated}, where $z=16$ corresponds to the in-focus plane.}
    %\fontsize{8.8pt}{6.5pt}\selectfont
    \vspace{-0.8em}
	\label{gene_w1}
	\centering
				\resizebox{0.9\linewidth}{!}{
					\begin{tabular}{l|c|c|c}
						\toprule
						&\multicolumn{1}{c}{z=01} & \multicolumn{1}{c}{z=05} & \multicolumn{1}{c}{z=13} \\
						\midrule
						GKMNet~\cite{quan2021gaussian}  & 31.50  & 33.43  & 32.78 \\
						NAFNet~\cite{chen2022simple}  & 32.37  & 34.43  & 36.33 \\
						MambaIRv2~\cite{guo2025mambairv2} &  30.63  & 33.25  & 36.29  \\
						SwinIR~\cite{liu_swin_2021} & 30.32   & 32.66  & 34.21\\
						Restormer~\cite{zamir2022restormer} & 32.18  & 33.79 & 36.33   \\
						MPT~\cite{zhang2024unified} & 32.87  & 34.72  & 35.56   \\
						\midrule
						DGNO-Cell & 32.41 & 34.87 & 36.25  \\
						DGNO-Face & \textbf{34.30} & \textbf{36.32} & \textbf{36.99}  \\
						\bottomrule
				\end{tabular}}
%			}
	% \vskip -0.1in
    \vspace{-1em}
\end{table}

\textbf{Validation on Downstream Tasks.}
Defocus blur can severely degrade cell detection performance~\cite{schmidt2018cell}, which is critical for many downstream biological analyses.
To evaluate the impact of defocus deblurring on cell detection, we apply StarDist~\cite{schmidt2018cell} to the BBBC006 dataset before and after restoration.
The results are reported in Table~\ref{tab:cell_detection_bbbc006} in terms of Average Precision (AP) under different Intersection-over-Union (IoU) thresholds, where higher AP indicates more accurately detected cells.
Compared with the blurred input, DGNO-based deblurring leads to substantial improvements in cell detection performance across all IoU thresholds.
In particular, DGNO-cell achieves the highest mean AP of 0.5561, while DGNO-Face attains the best performance at IoU threshold of 0.7211.
These results demonstrate that the proposed DGNO not only improves restoration quality but also better preserves cellular shape and boundary structures, thereby significantly benefiting downstream cell detection tasks.

\begin{table}[t]
\renewcommand\arraystretch{0.9}
	\centering
		\caption{Cell detection results on deblurred BBBC006.}
        \vspace{-0.8em}
		\label{tab:cell_detection_bbbc006}
		\setlength{\tabcolsep}{2pt}
		\resizebox{\linewidth}{!}{
%	\renewcommand\arraystretch{0.5}
%	\resizebox{0.9\linewidth}{!}{
		\begin{tabular}{lcccc}
			\toprule
			IoU & 0.5 & 0.7 & 0.9 & Mean AP \\
			\midrule
			blur
			& 0.5672 & 0.3084 & 0.0708 & 0.3154 \\
			\midrule
			MambaIRv2~\cite{guo2025mambairv2}
			&0.7972 & 0.6786 & 0.0939 &  0.5232 \\
			SwinIR~\cite{liu_swin_2021}
			&0.7914 &0.6722 &0.0643   &0.5160 \\
			Restormer~\cite{zamir2022restormer}
			&0.7910 & 0.6950 & 0.1122 &  0.5287\\
			MPT+EFCR~\cite{zhang2024unified}
			&0.7978 & 0.7026 & 0.1167  & 0.5350
			%		& \textcolor{blue}{0.7814} & \textcolor{blue}{0.6791} & \textcolor{blue}{0.2970} & \textcolor{blue}{0.5858} 
			\\
			DGNO-Face & 0.8055 & \textbf{0.7211} & 0.1355 &  0.5540 \\
			DGNO-Cell & \textbf{0.8070} & 0.7195 & \textbf{0.1418} &  \textbf{0.5561}
			\\
			
			\midrule
			sharp
			&0.8012 & 0.7175 & 0.2089  & 0.5758 \\
			\bottomrule
	\end{tabular}}
%}
\vspace{-2em}
\end{table}

\section{Conclusion}
We proposed DGNO, a novel operator learning framework for pathological image defocus
deblurring.
By incorporating discontinuous Galerkin principles into
neural operators, DGNO decomposes the global operator into element-local
volume operators and interface fluxes, enabling structured local
modeling and controlled cross-element interactions.
We developed both face-wise and cell-wise (P0DG) formulations of DGNO.
Experimental results on three microscopy defocus deblurring datasets and one natural image defocus deblurring dataset
show that DGNO consistently outperforms state-of-the-art methods in terms
of reconstruction accuracy and perceptual quality.

% Acknowledgements should only appear in the accepted version.
\section*{Acknowledgements}

This work was supported by the National Natural Science Foundation of China (Grant No. 62471182), Science and Technology Commission of Shanghai Municipality Basic Research Program (Grant No. 25JD1401300), Shanghai Rising-Star Program (Grant No. 24QA2702100), and the Science and Technology Commission of Shanghai Municipality (Grant No. 22DZ2229004)

\section*{Impact Statement}
This work proposes a physically interpretable neural operator framework for pathological defocus deblurring.
By modeling spatially varying blur as a piecewise integral operator, the proposed method may improve image restoration quality and support downstream microscopy analysis.
The approach is intended for research use and does not directly perform clinical diagnosis.

% In the unusual situation where you want a paper to appear in the
% references without citing it in the main text, use \nocite
\nocite{langley00}

\bibliography{example_paper}

@inproceedings{liu_swin_2021,
  title={Swin transformer: Hierarchical vision transformer using shifted windows},
  author={Liu, Ze and Lin, Yutong and Cao, Yue and Hu, Han and Wei, Yixuan and Zhang, Zheng and Lin, Stephen and Guo, Baining},
  booktitle={Proceedings of the IEEE/CVF international conference on computer vision},
  pages={10012--10022},
  year={2021}
}

@article{quan2021gaussian,
  title={Gaussian kernel mixture network for single image defocus deblurring},
  author={Quan, Yuhui and Wu, Zicong and Ji, Hui},
  journal={Advances in Neural Information Processing Systems},
  volume={34},
  pages={20812--20824},
  year={2021}
}

@inproceedings{zamir2022restormer,
  title={Restormer: Efficient transformer for high-resolution image restoration},
  author={Zamir, Syed Waqas and Arora, Aditya and Khan, Salman and Hayat, Munawar and Khan, Fahad Shahbaz and Yang, Ming-Hsuan},
  booktitle={Proceedings of the IEEE/CVF conference on computer vision and pattern recognition},
  pages={5728--5739},
  year={2022}
}

@inproceedings{cho2021rethinking,
  title={Rethinking coarse-to-fine approach in single image deblurring},
  author={Cho, Sung-Jin and Ji, Seo-Won and Hong, Jun-Pyo and Jung, Seung-Won and Ko, Sung-Jea},
  booktitle={Proceedings of the IEEE/CVF international conference on computer vision},
  pages={4641--4650},
  year={2021}
}

@inproceedings{chen2022simple,
  title={Simple baselines for image restoration},
  author={Chen, Liangyu and Chu, Xiaojie and Zhang, Xiangyu and Sun, Jian},
  booktitle={European conference on computer vision},
  pages={17--33},
  year={2022},
  organization={Springer}
}

@inproceedings{zhang2024unified,
  title={A unified framework for microscopy defocus deblur with multi-pyramid transformer and contrastive learning},
  author={Zhang, Yuelin and Zheng, Pengyu and Yan, Wanquan and Fang, Chengyu and Cheng, Shing Shin},
  booktitle={Proceedings of the IEEE/CVF Conference on Computer Vision and Pattern Recognition},
  pages={11125--11136},
  year={2024}
}

@inproceedings{guo2024mambair,
  title={Mambair: A simple baseline for image restoration with state-space model},
  author={Guo, Hang and Li, Jinmin and Dai, Tao and Ouyang, Zhihao and Ren, Xudong and Xia, Shu-Tao},
  booktitle={European conference on computer vision},
  pages={222--241},
  year={2024},
  organization={Springer}
}

@article{ljosa2012annotated,
  title={Annotated high-throughput microscopy image sets for validation},
  author={Ljosa, Vebjorn and Sokolnicki, Katherine L and Carpenter, Anne E},
  journal={Nature methods},
  volume={9},
  number={7},
  pages={637},
  year={2012}
}

@article{geng2022cervical,
  title={Cervical cytopathology image refocusing via multi-scale attention features and domain normalization},
  author={Geng, Xiebo and Liu, Xiuli and Cheng, Shenghua and Zeng, Shaoqun},
  journal={Medical Image Analysis},
  volume={81},
  pages={102566},
  year={2022},
  publisher={Elsevier}
}

@inproceedings{abuolaim2020defocus,
  title={Defocus deblurring using dual-pixel data},
  author={Abuolaim, Abdullah and Brown, Michael S},
  booktitle={European conference on computer vision},
  pages={111--126},
  year={2020},
  organization={Springer}
}

@inproceedings{lee2021iterative,
  title={Iterative filter adaptive network for single image defocus deblurring},
  author={Lee, Junyong and Son, Hyeongseok and Rim, Jaesung and Cho, Sunghyun and Lee, Seungyong},
  booktitle={Proceedings of the IEEE/CVF conference on computer vision and pattern recognition},
  pages={2034--2042},
  year={2021}
}

@article{vaswani2017attention,
  title={Attention is all you need},
  author={Vaswani, Ashish and Shazeer, Noam and Parmar, Niki and Uszkoreit, Jakob and Jones, Llion and Gomez, Aidan N and Kaiser, {\L}ukasz and Polosukhin, Illia},
  journal={Advances in neural information processing systems},
  volume={30},
  year={2017}
}

@article{quan2024deep,
  title={Deep single image defocus deblurring via gaussian kernel mixture learning},
  author={Quan, Yuhui and Wu, Zicong and Xu, Ruotao and Ji, Hui},
  journal={IEEE Transactions on Pattern Analysis and Machine Intelligence},
  year={2024},
  publisher={IEEE}
}

@inproceedings{guo2025mambairv2,
  title={Mambairv2: Attentive state space restoration},
  author={Guo, Hang and Guo, Yong and Zha, Yaohua and Zhang, Yulun and Li, Wenbo and Dai, Tao and Xia, Shu-Tao and Li, Yawei},
  booktitle={Proceedings of the Computer Vision and Pattern Recognition Conference},
  pages={28124--28133},
  year={2025}
}

@inproceedings{shi2015just,
  title={Just noticeable defocus blur detection and estimation},
  author={Shi, Jianping and Xu, Li and Jia, Jiaya},
  booktitle={Proceedings of the IEEE Conference on Computer Vision and Pattern Recognition},
  pages={657--665},
  year={2015}
}

@article{karaali2017edge,
  title={Edge-based defocus blur estimation with adaptive scale selection},
  author={Karaali, Ali and Jung, Claudio Rosito},
  journal={IEEE Transactions on Image Processing},
  volume={27},
  number={3},
  pages={1126--1137},
  year={2017},
  publisher={IEEE}
}

@incollection{yuan2007image,
  title={Image deblurring with blurred/noisy image pairs},
  author={Yuan, Lu and Sun, Jian and Quan, Long and Shum, Heung-Yeung},
  booktitle={ACM SIGGRAPH 2007 papers},
  pages={1--es},
  year={2007}
}

@article{zhao2019defocus,
  title={Defocus blur detection via multi-stream bottom-top-bottom network},
  author={Zhao, Wenda and Zhao, Fan and Wang, Dong and Lu, Huchuan},
  journal={IEEE transactions on pattern analysis and machine intelligence},
  volume={42},
  number={8},
  pages={1884--1897},
  year={2019},
  publisher={IEEE}
}

@inproceedings{nan2020deep,
  title={Deep learning for handling kernel/model uncertainty in image deconvolution},
  author={Nan, Yuesong and Ji, Hui},
  booktitle={Proceedings of the IEEE/CVF conference on computer vision and pattern recognition},
  pages={2388--2397},
  year={2020}
}

@article{ren2018deep,
  title={Deep non-blind deconvolution via generalized low-rank approximation},
  author={Ren, Wenqi and Zhang, Jiawei and Ma, Lin and Pan, Jinshan and Cao, Xiaochun and Zuo, Wangmeng and Liu, Wei and Yang, Ming-Hsuan},
  journal={Advances in neural information processing systems},
  volume={31},
  year={2018}
}

@article{yuan2008progressive,
  title={Progressive inter-scale and intra-scale non-blind image deconvolution},
  author={Yuan, Lu and Sun, Jian and Quan, Long and Shum, Heung-Yeung},
  journal={Acm Transactions on Graphics (TOG)},
  volume={27},
  number={3},
  pages={1--10},
  year={2008},
  publisher={ACM New York, NY, USA}
}

@article{li2020fourier,
  title={Fourier neural operator for parametric partial differential equations},
  author={Li, Zongyi and Kovachki, Nikola and Azizzadenesheli, Kamyar and Liu, Burigede and Bhattacharya, Kaushik and Stuart, Andrew and Anandkumar, Anima},
  journal={arXiv preprint arXiv:2010.08895},
  year={2020}
}

@article{kovachki2023neural,
  title={Neural operator: Learning maps between function spaces with applications to pdes},
  author={Kovachki, Nikola and Li, Zongyi and Liu, Burigede and Azizzadenesheli, Kamyar and Bhattacharya, Kaushik and Stuart, Andrew and Anandkumar, Anima},
  journal={Journal of Machine Learning Research},
  volume={24},
  number={89},
  pages={1--97},
  year={2023}
}

@article{lu2021learning,
  title={Learning nonlinear operators via DeepONet based on the universal approximation theorem of operators},
  author={Lu, Lu and Jin, Pengzhan and Pang, Guofei and Zhang, Zhongqiang and Karniadakis, George Em},
  journal={Nature machine intelligence},
  volume={3},
  number={3},
  pages={218--229},
  year={2021},
  publisher={Nature Publishing Group UK London}
}

@inproceedings{wei2023super,
  title={Super-resolution neural operator},
  author={Wei, Min and Zhang, Xuesong},
  booktitle={Proceedings of the IEEE/CVF Conference on Computer Vision and Pattern Recognition},
  pages={18247--18256},
  year={2023}
}

@inproceedings{liu2025difffno,
  title={DiffFNO: Diffusion Fourier Neural Operator},
  author={Liu, Xiaoyi and Tang, Hao},
  booktitle={Proceedings of the Computer Vision and Pattern Recognition Conference},
  pages={150--160},
  year={2025}
}

@book{goodman2005introduction,
  title={Introduction to Fourier optics},
  author={Goodman, Joseph W},
  year={2005},
  publisher={Roberts and Company publishers}
}

@book{hesthaven2008nodal,
  title={Nodal discontinuous Galerkin methods: algorithms, analysis, and applications},
  author={Hesthaven, Jan S and Warburton, Tim},
  year={2008},
  publisher={Springer}
}

@article{cui2023image,
  title={Image restoration via frequency selection},
  author={Cui, Yuning and Ren, Wenqi and Cao, Xiaochun and Knoll, Alois},
  journal={IEEE Transactions on Pattern Analysis and Machine Intelligence},
  volume={46},
  number={2},
  pages={1093--1108},
  year={2023},
  publisher={IEEE}
}

@inproceedings{zhang2022benchmarking,
  title={Benchmarking the robustness of deep neural networks to common corruptions in digital pathology},
  author={Zhang, Yunlong and Sun, Yuxuan and Li, Honglin and Zheng, Sunyi and Zhu, Chenglu and Yang, Lin},
  booktitle={International conference on medical image computing and computer-assisted intervention},
  pages={242--252},
  year={2022},
  organization={Springer}
}

@inproceedings{keaton2023celltranspose,
  title={Celltranspose: Few-shot domain adaptation for cellular instance segmentation},
  author={Keaton, Matthew R and Zaveri, Ram J and Doretto, Gianfranco},
  booktitle={Proceedings of the IEEE/CVF winter conference on applications of computer vision},
  pages={455--466},
  year={2023}
}

@inproceedings{schmidt2018cell,
  title={Cell detection with star-convex polygons},
  author={Schmidt, Uwe and Weigert, Martin and Broaddus, Coleman and Myers, Gene},
  booktitle={International conference on medical image computing and computer-assisted intervention},
  pages={265--273},
  year={2018},
  organization={Springer}
}

@inproceedings{nah2017deep,
  title={Deep multi-scale convolutional neural network for dynamic scene deblurring},
  author={Nah, Seungjun and Hyun Kim, Tae and Mu Lee, Kyoung},
  booktitle={Proceedings of the IEEE conference on computer vision and pattern recognition},
  pages={3883--3891},
  year={2017}
}

@inproceedings{tao2018scale,
  title={Scale-recurrent network for deep image deblurring},
  author={Tao, Xin and Gao, Hongyun and Shen, Xiaoyong and Wang, Jue and Jia, Jiaya},
  booktitle={Proceedings of the IEEE conference on computer vision and pattern recognition},
  pages={8174--8182},
  year={2018}
}

@inproceedings{gu2021interpreting,
  title={Interpreting super-resolution networks with local attribution maps},
  author={Gu, Jinjin and Dong, Chao},
  booktitle={Proceedings of the IEEE/CVF conference on computer vision and pattern recognition},
  pages={9199--9208},
  year={2021}
}

@article{cui2024revitalizing,
  title={Revitalizing convolutional network for image restoration},
  author={Cui, Yuning and Ren, Wenqi and Cao, Xiaochun and Knoll, Alois},
  journal={IEEE Transactions on Pattern Analysis and Machine Intelligence},
  volume={46},
  number={12},
  pages={9423--9438},
  year={2024},
  publisher={IEEE}
}

@article{kohli2025ring,
  title={Ring deconvolution microscopy: exploiting symmetry for efficient spatially varying aberration correction},
  author={Kohli, Amit and Angelopoulos, Anastasios N and McAllister, David and Whang, Esther and You, Sixian and Yanny, Kyrollos and Gasparoli, Federico M and Chang, Bo-Jui and Fiolka, Reto and Waller, Laura},
  journal={Nature methods},
  volume={22},
  number={6},
  pages={1311--1320},
  year={2025},
  publisher={Nature Publishing Group US New York}
}

@inproceedings{lin2025eamamba,
  title={Eamamba: Efficient all-around vision state space model for image restoration},
  author={Lin, Yu-Cheng and Xu, Yu-Syuan and Chen, Hao-Wei and Kuo, Hsien-Kai and Lee, Chun-Yi},
  booktitle={Proceedings of the IEEE/CVF International Conference on Computer Vision},
  pages={11708--11719},
  year={2025}
}
\bibliographystyle{icml2026}

%%%%%%%%%%%%%%%%%%%%%%%%%%%%%%%%%%%%%%%%%%%%%%%%%%%%%%%%%%%%%%%%%%%%%%%%%%%%%%%
%%%%%%%%%%%%%%%%%%%%%%%%%%%%%%%%%%%%%%%%%%%%%%%%%%%%%%%%%%%%%%%%%%%%%%%%%%%%%%%
% APPENDIX
%%%%%%%%%%%%%%%%%%%%%%%%%%%%%%%%%%%%%%%%%%%%%%%%%%%%%%%%%%%%%%%%%%%%%%%%%%%%%%%
%%%%%%%%%%%%%%%%%%%%%%%%%%%%%%%%%%%%%%%%%%%%%%%%%%%%%%%%%%%%%%%%%%%%%%%%%%%%%%%
\newpage
\appendix
\onecolumn

\section{Discontinuous Galerkin Weak Form: Volume and Interface Terms}
\label{sec:dg_weak_form}

Let $D \subset \mathbb{R}^d$ be a bounded domain and consider an operator equation
\begin{equation}
	\mathcal{L}(u) = f \quad \text{in } D,
	\label{eq:operator_equation}
\end{equation}
where $\mathcal{L}$ denotes a (possibly nonlinear) differential operator.
The domain is partitioned into non-overlapping elements
$\{D_e\}_{e=1}^E$ such that $D=\bigcup_{e=1}^E D_e$.
In the discontinuous Galerkin (DG) setting, trial and test functions are allowed
to be discontinuous across element interfaces.
Multiplying Eq.~\eqref{eq:operator_equation} by a test function $\phi$ and
integrating over an element $D_e$ yields the elementwise weak statement
\begin{equation}
	\int_{D_e} \phi\, \mathcal{L}(u)\, dx
	=
	\int_{D_e} \phi\, f\, dx.
	\label{eq:dg_element_test}
\end{equation}

\paragraph{Integration by parts and volume--surface decomposition.}
For clarity, we consider operators that admit a conservative form
$\mathcal{L}(u)=\nabla\!\cdot\!\mathbf{F}(u)$.
The product rule for the divergence operator gives
\begin{equation}
	\nabla\cdot(\phi\,\mathbf{F}(u))
	=
	\nabla\phi\cdot\mathbf{F}(u)
	+
	\phi\,\nabla\cdot\mathbf{F}(u).
	\label{eq:product_rule}
\end{equation}
Integrating Eq.~\eqref{eq:product_rule} over $D_e$ and applying the divergence
theorem yields the integration-by-parts formula
\begin{equation}
	\int_{D_e}\phi\,\nabla\cdot\mathbf{F}(u)\,dx
	=
	-\int_{D_e}\nabla\phi\cdot\mathbf{F}(u)\,dx
	+
	\int_{\partial D_e}\phi\,\mathbf{F}(u)\cdot\mathbf{n}\,ds,
	\label{eq:dg_ibp}
\end{equation}
where $\mathbf{n}$ denotes the outward unit normal on $\partial D_e$.
Substituting \eqref{eq:dg_ibp} into \eqref{eq:dg_element_test} leads to the
weak formulation
\begin{equation}
	-\int_{D_e}\nabla\phi\cdot\mathbf{F}(u)\,dx
	+
	\int_{\partial D_e}\phi\,\mathbf{F}(u)\cdot\mathbf{n}\,ds
	=
	\int_{D_e}\phi\,f\,dx.
\end{equation}
The first term is an element-local \emph{volume contribution}, while the second
is a \emph{surface contribution} on $\partial D_e$.

\paragraph{Numerical flux and interface term.}
Since discontinuous Galerkin methods allow trial and test functions to be
discontinuous across element interfaces, the physical normal flux
$\mathbf{F}(u)\cdot \mathbf{n}$ is generally not uniquely defined on
$\partial D_e$.
Let $u^-$ and $u^+$ denote the interior and exterior traces of $u$ on an
interface.
To resolve this ambiguity, the physical flux is replaced by a
\emph{numerical flux} $\widehat{F}(u^-,u^+)$, which provides a consistent and
stable approximation of the inter-element flux.
Substituting the numerical flux into the boundary term yields the
discontinuous Galerkin weak formulation given below.

\paragraph{DG weak form.}
The resulting discontinuous Galerkin weak formulation on each element $D_e$
consists of an element-local volume term and an interface (flux) term:
\begin{equation}
		-\int_{D_e} \nabla \phi \cdot \mathbf{F}(u)\, dx
		\;+\;
		\int_{\partial D_e} \phi\, \widehat{F}(u^-,u^+)\, ds
		=
		\int_{D_e} \phi\, f\, dx.
	\label{eq:dg_weak_form}
\end{equation}
The numerical flux $\widehat{F}$ constitutes the sole mechanism for
inter-element coupling and plays a central role in ensuring stability and
consistency of DG discretizations.
Common choices of $\widehat{F}$ include the central flux, upwind-type fluxes,
and interior-penalty fluxes, each leading to a different DG scheme with distinct
stability and dissipation properties.
The central flux is given by
\begin{equation}
	\widehat{F}_{\mathrm{cen}}(u^-,u^+)
	=
	\{\mathbf F(u)\}\cdot \mathbf n
	=
	\frac{1}{2}\big(\mathbf F(u^-)+\mathbf F(u^+)\big)\cdot \mathbf n,
\end{equation}
and the upwind-type (Rusanov/LLF) flux
\begin{equation}
	\widehat{F}_{\mathrm{LLF}}(u^-,u^+)
	=
	\{\mathbf F(u)\}\cdot \mathbf n
	-
	\frac{\alpha_e}{2}[u],
	\qquad [u]=u^+-u^-,
\end{equation}
where $\alpha_e$ denotes a local upper bound on the characteristic wave speed.
For diffusion operators of the form $-\nabla\cdot(\kappa\nabla u)$, the symmetric
interior-penalty Galerkin (SIPG) fluxes are commonly used:
\begin{equation}
	\widehat{u} = \{u\}, \qquad
	\widehat{(\kappa\nabla u)\cdot \mathbf n}
	=
	\{\kappa\nabla u\}\cdot \mathbf n
	-
	\tau\, [u],
\end{equation}
where $\{\cdot\}$ denotes the arithmetic average across the interface and
$\tau>0$ is a penalty parameter controlling the strength of inter-element
coupling.

\section{Relation to Classical Discontinuous Galerkin Methods}

Classical DG methods are developed for the numerical discretization of PDEs in differential form. As shown in Appendix~A, the DG weak formulation decomposes into an element-local volume term and an interface flux term, arising directly from integration by parts applied to the differential operator, where the numerical flux $\widehat{F}(u^-, u^+)$ resolves the non-uniqueness of the physical flux at element interfaces. DGNO, by contrast, is not derived from a differential formulation. Each operator layer in Eq.~(3) is defined as a kernel integral operator over the entire domain $D$, which does not admit a natural integration-by-parts structure, so no interface terms arise at the continuous operator level.

DGNO instead introduces discontinuities at the level of operator discretization. The global kernel integral is restricted to non-overlapping elements via domain partitioning $D = \bigcup_{e=1}^{E} D_e$, yielding element-local operators as in Eq.~(7). Since each local operator is a truncated restriction of the global one, the cross-element interactions encoded in the original integral are lost and must be explicitly restored through operator-valued numerical fluxes at element interfaces, constructed from the learned boundary operator matrices $\mathbf{K}_e^{\partial,f}$. The essential difference from classical DG is therefore that the volume--interface decomposition in DGNO is a deliberate discretization strategy for restoring cross-element coupling, rather than a necessary consequence of integration by parts. Despite this difference in origin, DGNO inherits the modularity, locality, and controlled global coupling of classical DG methods, and remains compatible with the neural operator learning framework.

\section{Operator-valued Numerical Fluxes.}
\label{app:Fluxes}
The operator-valued numerical flux
$\widehat{\mathcal{F}}(\cdot,\cdot)$ determines how interface information
from neighboring elements is combined to restore inter-element coupling.
We consider several standard choices inspired by classical DG
discretizations, adapted to the operator-level setting.

\emph{Central flux} is defined by averaging the boundary operators from
the two sides of the interface,
\begin{equation}
	\widehat{\mathcal{F}}_{\mathrm{central}}
	\!\left(
	\mathbf{K}^{\partial,f}_{e},
	\mathbf{K}^{\partial,f}_{e'}
	\right)
	=
	\frac{1}{2}
	\left(
	\mathbf{K}^{\partial,f}_{e}
	+
	\mathbf{K}^{\partial,f}_{e'}
	\right),
	\label{eq:dgno_flux_central}
\end{equation}
which yields a symmetric coupling between adjacent elements.

\emph{Jump flux} penalizes the operator discontinuity across the
interface and corresponds to a symmetric interior penalty (SIP)–style
stabilization,
\begin{equation}
	\widehat{\mathcal{F}}_{\mathrm{jump}}
	\!\left(
	\mathbf{K}^{\partial,f}_{e},
	\mathbf{K}^{\partial,f}_{e'}
	\right)
	=
	-\tau
	\left(
	\mathbf{K}^{\partial,f}_{e}
	-
	\mathbf{K}^{\partial,f}_{e'}
	\right),
	\label{eq:dgno_flux_jump}
\end{equation}
where $\tau$ is a learnable penalty coefficient controlling the strength
of inter-element coupling.
Combining consistency and stabilization, the \emph{Avg+Jump flux} takes
the form
\begin{equation}
	\begin{aligned}
		\widehat{\mathcal{F}}_{\mathrm{avg+jump}}
		\!\left(
		\mathbf{K}^{\partial,f}_{e},
		\mathbf{K}^{\partial,f}_{e'}
		\right)
		={}&
		\frac{1}{2}
		\left(
		\mathbf{K}^{\partial,f}_{e}
		+
		\mathbf{K}^{\partial,f}_{e'}
		\right) 
		&
		-
		\tau
		\left(
		\mathbf{K}^{\partial,f}_{e}
		-
		\mathbf{K}^{\partial,f}_{e'}
		\right),
	\end{aligned}
	\label{eq:dgno_flux_avg_jump}
\end{equation}

which is analogous to the classical SIP formulation and provides a
balance between accuracy and stability.

Finally, we consider an \emph{upwind-style flux} that introduces a
directional bias in operator space.
Specifically, we define a data-dependent weighting coefficient
\begin{equation}
	\alpha_{e,e'}^{f}
	=
	\sigma\!\left(
	s\!\left(\mathbf{K}^{\partial,f}_{e}\right)
	-
	s\!\left(\mathbf{K}^{\partial,f}_{e'}\right)
	\right),
	\label{eq:dgno_flux_upwind_alpha}
\end{equation}
where $s(\cdot)$ denotes a scalar summary statistic of the boundary
operator (e.g., the mean over matrix entries), and $\sigma(\cdot)$ is the
sigmoid function.
The upwind flux is then given by
\begin{equation}
	\widehat{\mathcal{F}}_{\mathrm{upwind}}
	=
	\alpha_{e,e'}^{f}\,\mathbf{K}^{\partial,f}_{e}
	+
	\bigl(1-\alpha_{e,e'}^{f}\bigr)\,\mathbf{K}^{\partial,f}_{e'},
	\label{eq:dgno_flux_upwind}
\end{equation}
which adaptively selects the dominant operator contribution across the
interface.

\section{Boundary Conditions}
\label{app:boundary_conditions}

Boundary conditions in the proposed DG neural operator are incorporated at the
\emph{operator level} through the operator-valued numerical flux
$\widehat{\mathcal{F}}(\cdot,\cdot)$.
Rather than imposing constraints directly on the latent representations or
solution values, boundary conditions are enforced by modifying the operator-level flux evaluation on boundary faces.
This treatment follows the discontinuous Galerkin philosophy and enables a
unified handling of interior interfaces and physical boundaries within the same
operator framework.
For a boundary face $f \subset \partial D_e \cap \partial D$, the operator-valued numerical flux is evaluated by appropriately replacing the exterior operator argument according to the prescribed boundary condition.

\paragraph{Dirichlet boundary conditions.}
For Dirichlet boundary conditions, the exterior operator
contribution is replaced by a boundary operator constructed from the given
boundary data, denoted by $\mathbf{K}^{\partial,f}_{\mathrm{bc}}$.
Specifically, we set
$
\mathbf{K}^{\partial,f}_{\mathrm{bc}}
=
\mathbf{K}^{\partial,f}_{e}.
$
The numerical flux on the boundary face is then evaluated as
$
\widehat{\mathcal{F}}
\bigl(
\mathbf{K}^{\partial,f}_{e},
\mathbf{K}^{\partial,f}_{\mathrm{bc}}
\bigr)
=
\widehat{\mathcal{F}}
\bigl(
\mathbf{K}^{\partial,f}_{e},
\mathbf{K}^{\partial,f}_{e}
\bigr).
$

\paragraph{Neumann boundary conditions.}
For Neumann boundary conditions, the prescribed normal flux on the physical
boundary is incorporated directly through the operator-valued numerical flux.
In the case of homogeneous Neumann conditions (zero normal flux), the exterior
operator contribution is set to zero, i.e.,
$\mathbf{K}^{\partial,f}_{\mathrm{bc}} = \mathbf{0}$,
and the numerical flux on a boundary face
$f \subset \partial D_e \cap \partial D$ is evaluated as
$
\widehat{\mathcal{F}}
\bigl(
\mathbf{K}^{\partial,f}_{e},
\mathbf{0}
\bigr).
$
This corresponds to enforcing a zero-flux condition at the operator
level.

\paragraph{Periodic boundary conditions.}
For periodic boundary conditions, boundary faces are paired and treated as
interior interfaces.
The boundary operators on paired faces are exchanged between the corresponding
elements, and the operator-valued numerical flux is evaluated in the same manner
as for interior faces.
This yields a seamless coupling across periodic boundaries at the operator
level.

\section{Additional Experiments and Analysis}
\label{app:additional_experiments}

\paragraph{Sensitivity to element size and operator iterations.}
We ablate DGNO-Cell over the element size $W_e$ and the number of
operator iterations $T$ on BBBC006$_{w1}$.
As shown in Table~\ref{tab:sensitivity_win_t}, the model is insensitive to these
hyperparameters across a broad range: for $W_e=4$--$16$ and
$T=1$--$4$, PSNR remains around 37 dB with only minor variation.
The best result is obtained at $W_e=8$ and $T=2$.
Clear degradation appears only when using a much larger element size ($W_e=32$), especially for larger $T$, indicating that the proposed DGNO-cell is robust to moderate changes in element granularity and operator depth.

\paragraph{Transfer beyond pathology deblurring.}
In addition to the non-pathology DPDD evaluation in Table~\ref{tab:dpdd}, we
further evaluate DGNO on the RealDOF test set~\cite{lee2021iterative}.
Table~\ref{tab:realdof} shows that DGNO achieves the best PSNR, SSIM, and LPIPS
among all compared methods, further demonstrating that the proposed
operator-level formulation transfers beyond pathology images.
Together with the DPDD results, this supports the broader applicability of DGNO
to spatially varying defocus deblurring in non-pathology domains.

\paragraph{Runtime, memory, and scaling.}
Complementing the performance--parameter and resolution--FLOPs analysis in Fig.~\ref{fig:performance}, Table~\ref{tab:runtime_memory_scaling} reports GPU memory usage and throughput across image sizes.
DGNO consistently uses substantially less memory and maintains higher throughput as the image resolution increases.
This confirms that the proposed element-wise operator decomposition provides a favorable runtime--memory trade-off and scales more efficiently to larger images.

\begin{table*}[t]
	\centering
	\scriptsize
	\renewcommand\arraystretch{0.95}
	\setlength{\tabcolsep}{2pt}
	\begin{minipage}[t]{0.28\textwidth}
		\centering
		\caption{Sensitivity to $W_e$ and $T$ on BBBC006$_{w1}$ (PSNR dB).}
		\label{tab:sensitivity_win_t}
		\resizebox{\linewidth}{!}{
			\begin{tabular}{c|cccc}
				\toprule
				$W_e \backslash T$ & 1 & 2 & 3 & 4 \\
				\midrule
				4  & 36.94 & 36.86 & 37.11 & 36.89 \\
				8  & 36.89 & \textbf{37.22} & 37.10 & 37.04 \\
				16 & 37.07 & 37.11 & 37.01 & 36.98 \\
				32 & 36.85 & 36.91 & 36.41 & 36.39 \\
				\bottomrule
			\end{tabular}
		}
	\end{minipage}
	\hfill
	\begin{minipage}[t]{0.4\textwidth}
		\centering
		\caption{Quantitative evaluation on RealDOF dataset~\cite{lee2021iterative}.}
		\label{tab:realdof}
		\resizebox{\linewidth}{!}{
			\begin{tabular}{l|ccc}
				\toprule
				Method & PSNR & SSIM & LPIPS \\
				\midrule
				DPDNet\cite{abuolaim2020defocus}  & 22.52 & 0.644 & 0.583 \\
				IFAN\cite{lee2021iterative}    & 24.71 & 0.748 & 0.304 \\
				EAMamba\cite{lin2025eamamba} & 24.59 & 0.761 & 0.309 \\
				\midrule
				DGNO (Ours) & \textbf{25.08} & \textbf{0.781} & \textbf{0.275} \\
				\bottomrule
			\end{tabular}
		}
	\end{minipage}
	\hfill
	\begin{minipage}[t]{0.28\textwidth}
		\centering
		\caption{Memory (GB) / throughput (img/s) versus image size.}
		\label{tab:runtime_memory_scaling}
		\resizebox{\linewidth}{!}{
			\begin{tabular}{c|cccc}
				\toprule
				Size & SwinIR & Restormer & MPT & DGNO \\
				\midrule
				128 & 4/29  & 3/24  & 2/7 & \textbf{1/30} \\
				256 & 16/5  & 12/11 & 10/9 & \textbf{4/27} \\
				384 & 36/2  & 27/5  & 24/3 & \textbf{9/18} \\
				512 & 63/1  & 48/3  & 41/2 & \textbf{15/10} \\
				\bottomrule
			\end{tabular}
		}
	\end{minipage}
\end{table*}

\paragraph{Penalty coefficient analysis.}
Table~\ref{Penalty} evaluates the penalty coefficient in the numerical flux.
The learnable coefficient achieves the best performance for both DGNO-Face and DGNO-Cell, showing that adaptive flux strength is beneficial.

\paragraph{Generalization across defocus levels.}
Table~\ref{gene_w2} reports generalization on BBBC006$_{w2}$ under decreasing defocus levels.
DGNO remains consistently competitive across all focal planes, confirming its robustness to different blur severities.

\begin{table*}[!htb]
	\centering
	\small
	\renewcommand\arraystretch{0.9}
	\setlength{\tabcolsep}{2pt}
	\begin{minipage}[!htb]{0.42\textwidth}
		\centering
		\begin{sc}
			\caption{Ablation study of the penalty coefficient in the numerical flux on BBBC006$_{w1}$.}
			\label{Penalty}
			\begin{tabular}{l|cc|cc}
				\toprule
				Penalty &\multicolumn{2}{c}{DGNO-Face} & \multicolumn{2}{c}{DGNO-Cell} \\
				coefficient& PSNR & SSIM & PSNR & SSIM \\
				\midrule
				0.25       & 36.93 & 0.958 & 36.96 & 0.958  \\
				0.5        & 37.05 & 0.958 & 36.80 & 0.957  \\
				1          & 37.03 & 0.958 & 36.90 & 0.957  \\
				learnable  & \textbf{37.09} & \textbf{0.958} & \textbf{37.22} & \textbf{0.959}  \\
				\bottomrule
			\end{tabular}
		\end{sc}
	\end{minipage}
	\hfill
	\begin{minipage}[!htb]{0.54\textwidth}
		\centering
		\caption{Generalization performance of different algorithms under decreasing levels of defocus blur ($z=01, 05, 13$) on BBBC006$_{w2}$~\cite{ljosa2012annotated}, where $z=16$ corresponds to the in-focus plane.}
		\label{gene_w2}
		\begin{sc}
			\begin{tabular}{l|c|c|c}
				\toprule
				&\multicolumn{1}{c}{z=01} & \multicolumn{1}{c}{z=05} & \multicolumn{1}{c}{z=13} \\
				Method& PSNR  & PSNR  & PSNR  \\
				\midrule
				NAFNet~\cite{chen2022simple}  & 27.45  & 29.26  & 31.46 \\
				MambaIRv2~\cite{guo2025mambairv2} &  27.43  & 29.87  & 31.37  \\
				SwinIR~\cite{liu_swin_2021} & 25.25   & 27.13  & 28.28\\
				Restormer~\cite{zamir2022restormer} & 27.30  & 28.75 & 32.55   \\
				MPT~\cite{zhang2024unified} & 25.75  & 27.24  & 32.44   \\
				\midrule
				DGNO-Cell & 29.69 & 31.54 & \textbf{32.95}  \\
				DGNO-Face & \textbf{29.72} & \textbf{31.61} & 32.77  \\
				\bottomrule
			\end{tabular}
		\end{sc}
	\end{minipage}
	\vspace{-0.5em}
\end{table*}

\begin{figure}[!htb]
	\centering
	\begin{subfigure}[t]{0.48\textwidth}
		\centering
		\includegraphics[width=\textwidth]{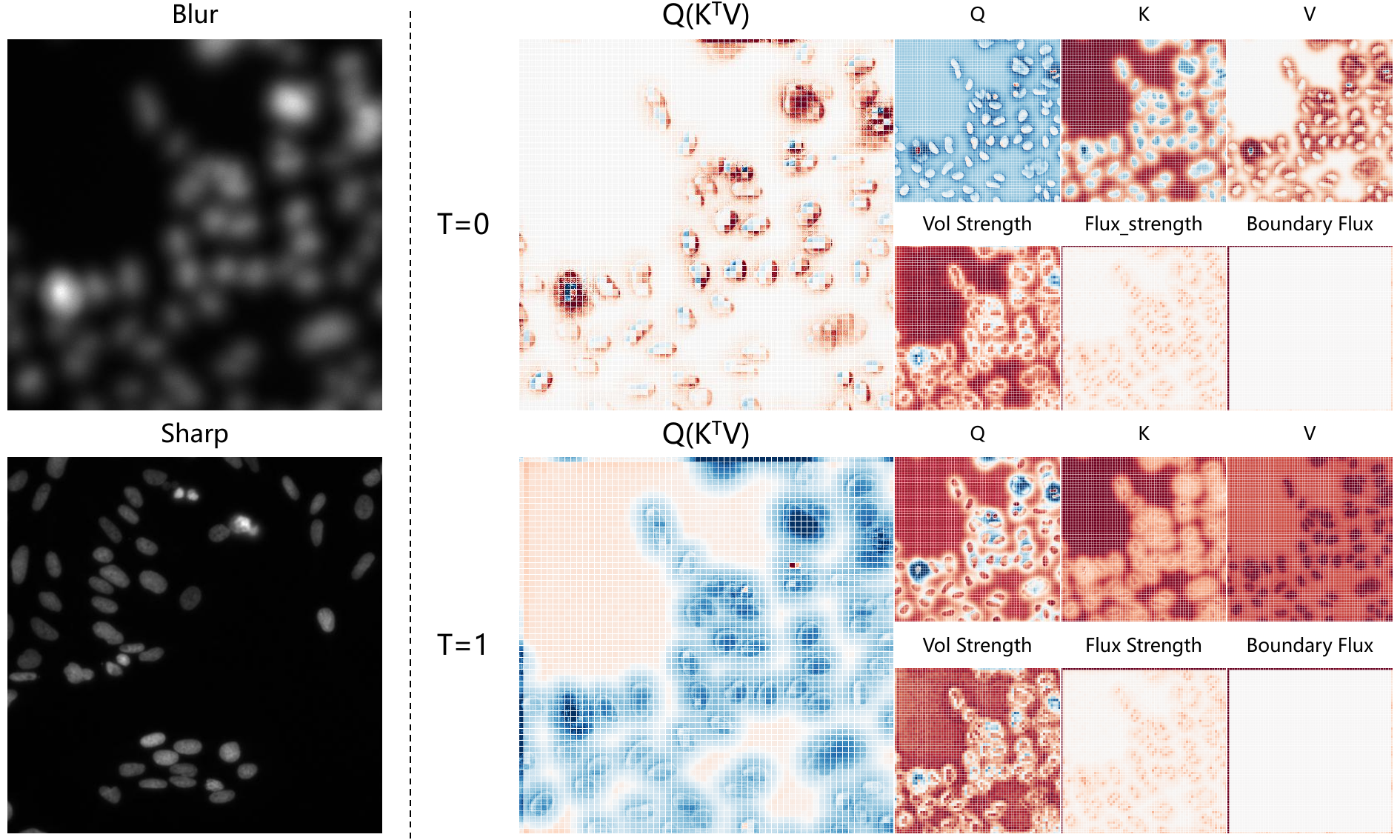}
		%		\caption{Visualization of operator evolution and DG flux behavior.}
		%		\label{fig:e3}
	\end{subfigure}
	\hfill
	\begin{subfigure}[t]{0.48\textwidth}
		\centering
		\includegraphics[width=\textwidth]{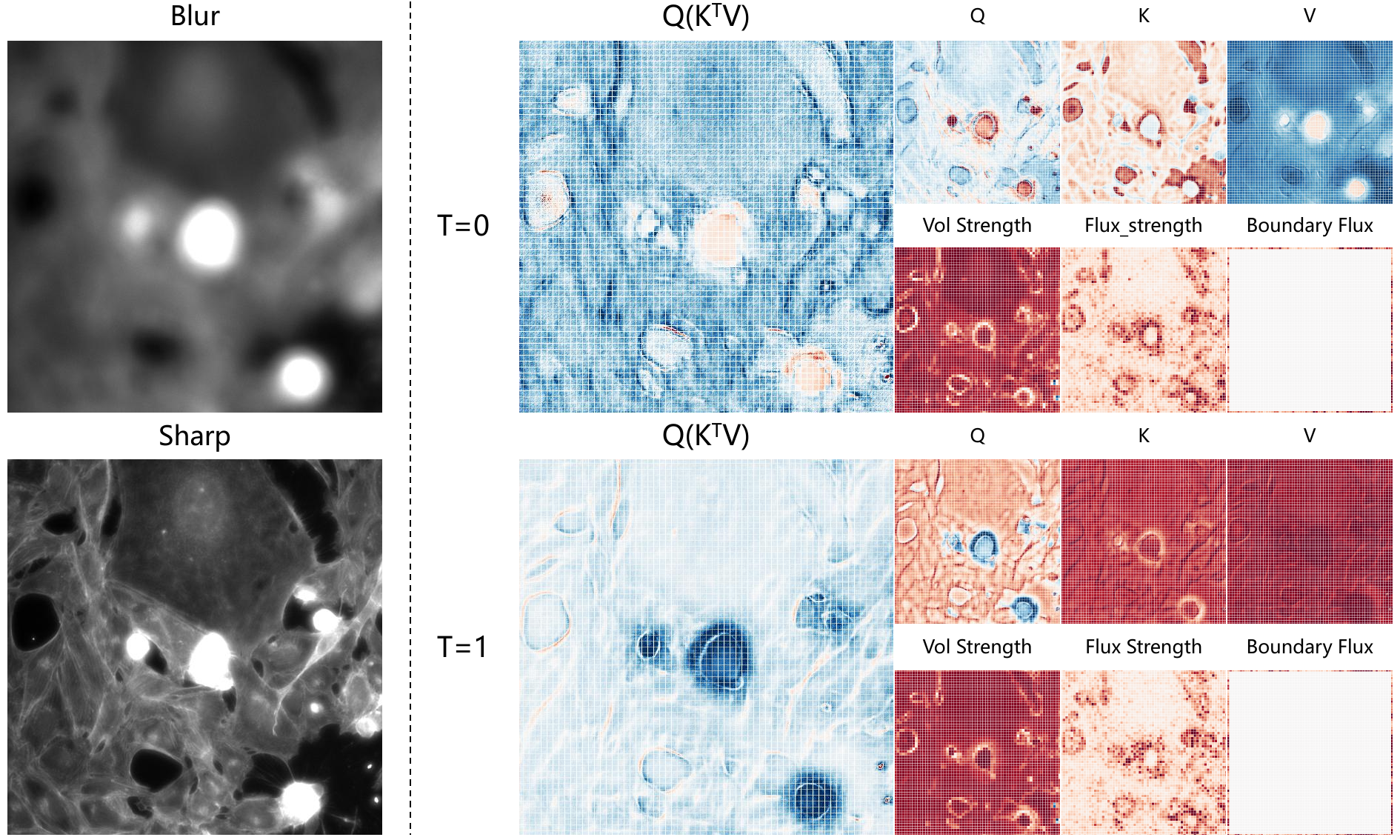}
		%		\caption{Visualization of the learned dynamic basis functions, latent representations $Q(K^{\top}V)$, and DG flux behavior on BBBC006$_{w1}$.}
		%		\label{fig:e4}
	\end{subfigure}
	\caption{Visualization of the learned dynamic basis functions, latent representations $Q(K^{\top}V)$, and DG flux and boundary behavior on BBBC006~\cite{ljosa2012annotated}.}
	\label{fig:qualitative}
	\vspace{-10pt}
\end{figure}

\paragraph{Qualitative analysis of learned operators.}
Figure~\ref{fig:qualitative} visualizes the internal operator behavior of DGNO on BBBC006~\cite{ljosa2012annotated}. 
The learned dynamic basis functions show spatially adaptive responses, indicating that DGNO captures local blur variations rather than relying on a fixed global operator. 
The latent representations $Q(K^{\top}V)$ highlight restoration-relevant cell structures and blurred boundaries. 
Moreover, the DG flux and boundary responses are concentrated around element interfaces and structural transitions, showing that the numerical flux effectively couples neighboring elements while preserving local spatial heterogeneity.

\section{Dataset Description.}
\label{appendix: datasets}
\noindent\textbf{BBBC006}~\cite{ljosa2012annotated}: This dataset from the Broad Bioimage Benchmark Collection (BBBC) consists of fluorescence microscopy images in two channels, denoted as $w1$ (Hoechst-stained nuclei) and $w2$ (Phalloidin-stained actin). 
Following the official protocol, images captured at the optimal focal plane ($z$-stack = 16) are used as sharp ground-truth references, while those at defocused planes ($z$-stack = [2, 6, 10]) serve as blurry inputs for training. 
All images are single-channel grayscale with a resolution of $696\times520$. 
A total of 6,144 image pairs are used, with a training/testing split of 4:1.

\vspace{0.5em}
\noindent\textbf{3DHistech}~\cite{geng2022cervical}: This cytopathology dataset is captured using a 3DHistech digital scanner across multiple focal planes. 
For each cell sample, the image containing the maximum number of cells in focus is regarded as the sharp ground truth. 
The dataset contains 94,973 image patches of size $256\times256$, divided into 66,976 for training, 9,088 for validation, and 18,909 for testing.

\section{Training Objective}
\label{app:training_objective}

To encourage restoration consistency in both spatial and frequency domains, we
employ a combined multi-scale reconstruction loss:
\begin{equation}
	\begin{aligned}
		L_{\mathrm{spatial}}
		&=
		\sum_{s=1}^{3}
		\frac{1}{E_s}
		\left\| \hat{Y}_s - Y_s \right\|_1, \\
		L_{\mathrm{frequency}}
		&=
		\sum_{s=1}^{3}
		\frac{1}{E_s}
		\left\| \mathcal{F}(\hat{Y}_s) - \mathcal{F}(Y_s) \right\|_1, \\
		L
		&=
		L_{\mathrm{spatial}} + \lambda L_{\mathrm{frequency}},
		\quad \lambda = 0.1.
	\end{aligned}
	\label{eq:loss-final}
\end{equation}
Here, $s$ denotes the scale index, $\mathcal{F}$ is the fast Fourier transform
(FFT), $E_s$ is a normalization factor, and $\hat{Y}_s$ and $Y_s$ denote the
restored output and target image at scale $s$, respectively.
This objective balances spatial-domain fidelity and frequency-domain consistency
across scales.

\begin{figure*}[!htb]
	\centering
	\begin{center}
		\includegraphics[width=1\textwidth]{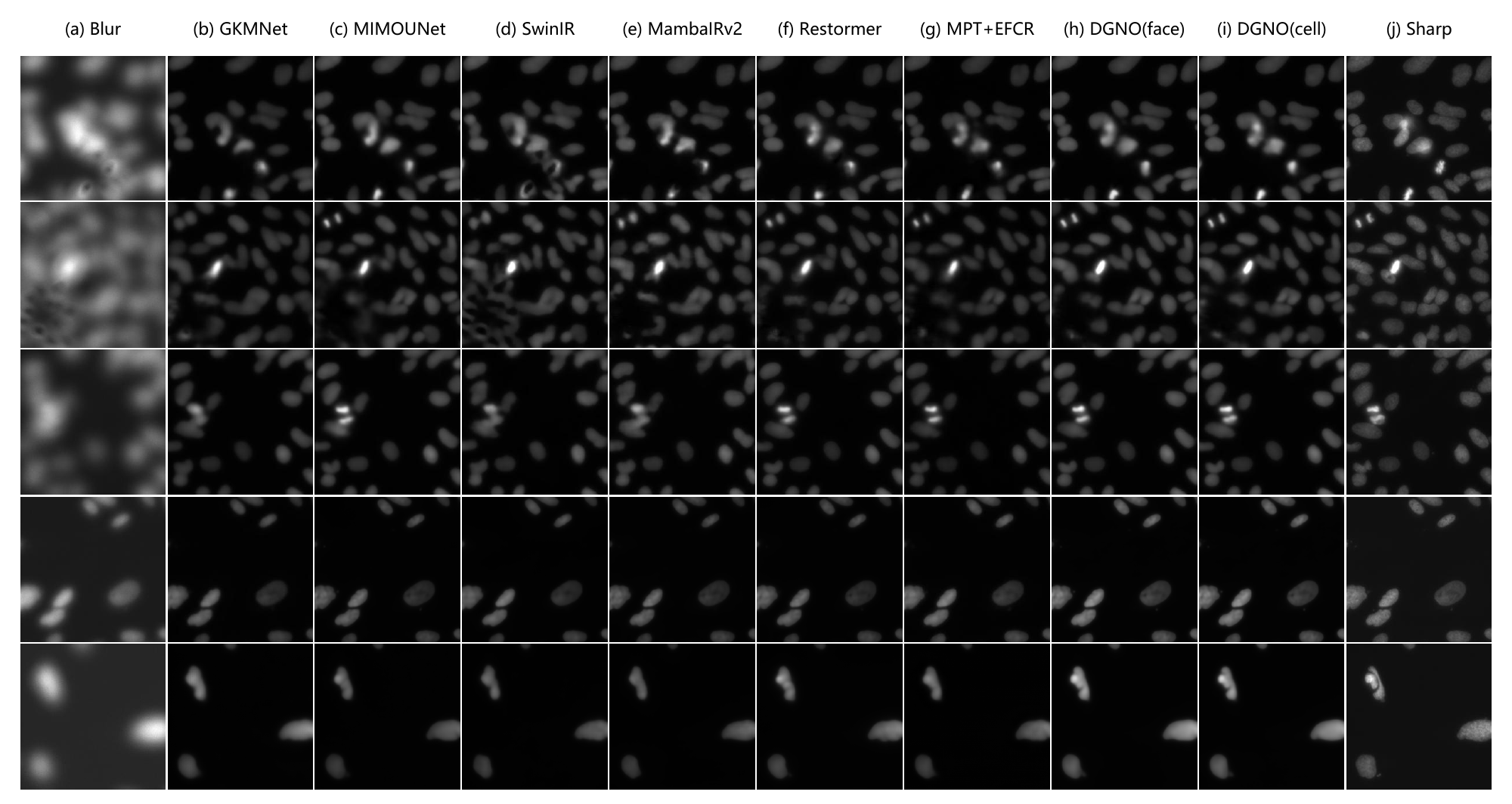}
		\caption{
			Visual comparison of single image defocus deblur approaches on BBBC006$_{w1}$~\cite{ljosa2012annotated}}
		\label{fig:Visual_w1}
	\end{center}
	\vspace{-15pt}
\end{figure*}

\begin{figure*}[!htb]
	\centering
	\begin{center}
		\includegraphics[width=1\textwidth]{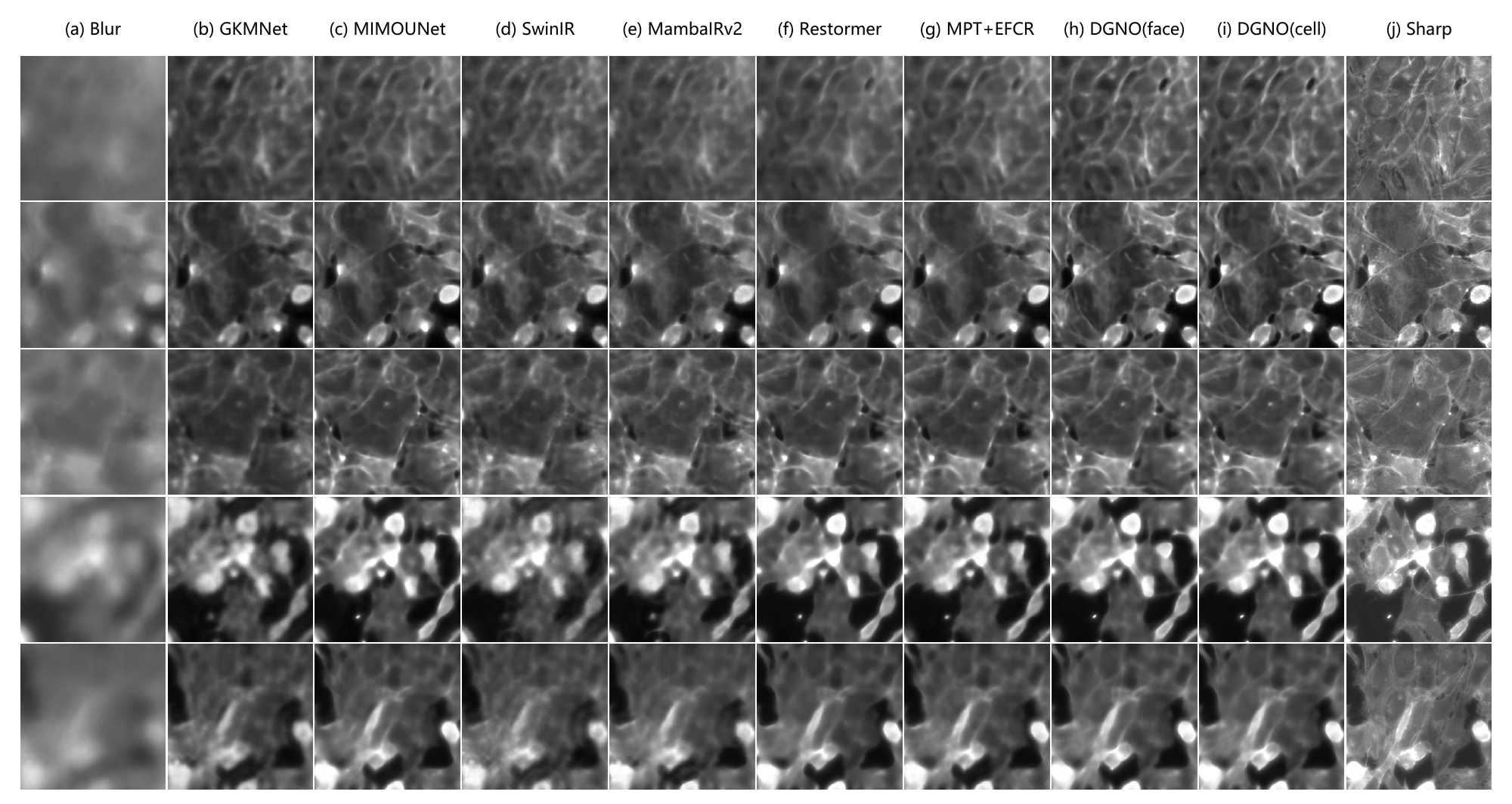}
		\caption{
			Visual comparison of single image defocus deblur approaches on BBBC006$_{w2}$~\cite{ljosa2012annotated}}
		\label{fig:Visual_w2}
	\end{center}
	\vspace{-15pt}
\end{figure*}

\begin{figure*}[!htb]
	\centering
	\begin{center}
		\includegraphics[width=1\textwidth]{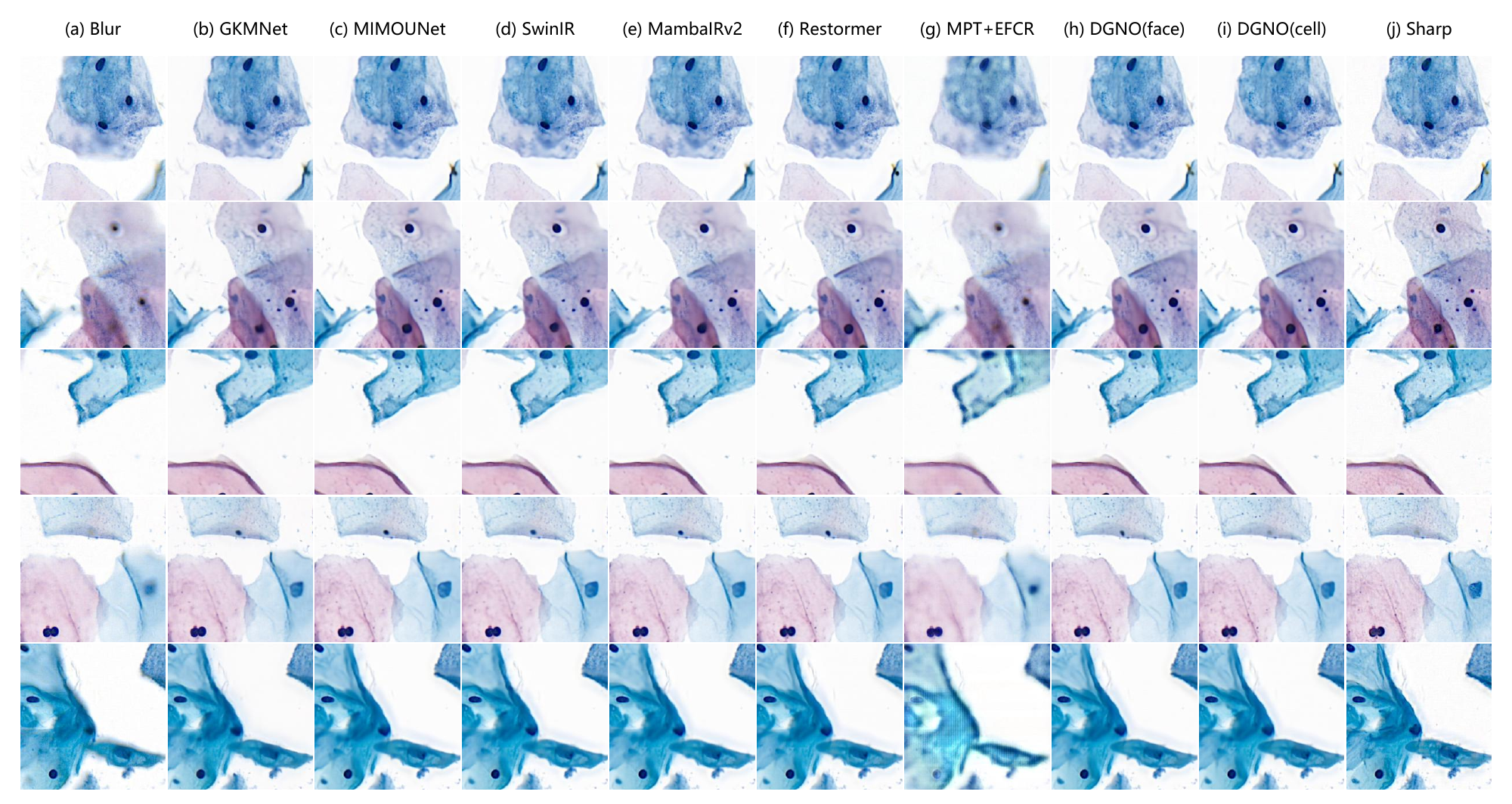}
		\caption{
			Visual comparison of single image defocus deblur approaches on 3DHistech~\cite{geng2022cervical}}
		\label{fig:Visual_3DHistech}
	\end{center}
	\vspace{-15pt}
\end{figure*}

% \begin{table}[t]
% 	\centering
% 	\caption{Comparison of modeling paradigms for defocus deblurring$\checkmark$: explicitly supported; 
% 		$\triangle$: partially or implicitly supported; 
% 		$\times$: not explicitly modeled.}
% 	\label{tab:modeling_comparison}
% 	\renewcommand{\arraystretch}{1.2}
% 	\resizebox{\linewidth}{!}{
% 	\begin{tabular}{lcccc}
% 		\toprule
% 		\textbf{Modeling Paradigm} 
% 		& \textbf{Physical Interpretability} 
% 		& \textbf{Explicit Operator / Kernel} 
% 		& \textbf{Spatially Varying Blur} 
% 		& \textbf{Inductive Bias} \\
% 		\midrule
% 		Convolutional Networks (CNNs) 
% 		& $\triangle$ (implicit PSF) 
% 		& $\times$ (fixed convolution) 
% 		& $\times$ (global stationarity) 
% 		& Convolutional prior \\
		
% 		Attention-based Models (Transformers) 
% 		& $\times$ (data-driven) 
% 		& $\times$ (no explicit kernel) 
% 		& $\triangle$ (position-aware) 
% 		& Global token mixing \\
		
% 		State-space Models (Mamba) 
% 		& $\times$ (non-optical dynamics) 
% 		& $\times$ (latent transitions) 
% 		& $\triangle$ (implicit adaptation) 
% 		& Sequential modeling \\
		
% 		\textbf{Neural Operator-based Models (Ours)} 
% 		& \textbf{\checkmark} (integral operator) 
% 		& \textbf{\checkmark} (learned integral kernels) 
% 		& \textbf{\checkmark} (space-variant PSF) 
% 		& \textbf{Physics-inspired operator prior} \\
% 		\bottomrule
% 	\end{tabular}
% }
% \end{table}

%%%%%%%%%%%%%%%%%%%%%%%%%%%%%%%%%%%%%%%%%%%%%%%%%%%%%%%%%%%%%%%%%%%%%%%%%%%%%%%
%%%%%%%%%%%%%%%%%%%%%%%%%%%%%%%%%%%%%%%%%%%%%%%%%%%%%%%%%%%%%%%%%%%%%%%%%%%%%%%

\end{document}